\documentclass[australian,english,prl, singlespace, twocolumn]{revtex4-1}
\usepackage[T1]{fontenc}
\usepackage[latin9]{inputenc}
\usepackage{xcolor}
\usepackage{verbatim}
\usepackage{amsmath}
\usepackage{amssymb}
\usepackage{graphicx}

\makeatletter

\providecommand{\tabularnewline}{\\}

\@ifundefined{definecolor}
 {\usepackage{color}}{}
\makeatother

\makeatother

\usepackage{babel}
\begin{document}
\title{Wigner's Friend paradoxes: consistency with weak-contextual and weak-macroscopic
realism models}
\author{Ria Joseph, Manushan Thenabadu, Channa Hatharasinghe, Jesse Fulton,
Run-Yan Teh, P. D. Drummond and M. D. Reid $^{1}$}
\affiliation{$^{1}$ Centre for Quantum Science and Technology Theory, Swinburne
University of Technology, Melbourne 3122, Australia}
\begin{abstract}
Wigner's friend paradoxes highlight contradictions between measurements
made by Friends inside a laboratory and superobservers outside a laboratory,
who have access to an entangled state of the measurement apparatus.
The contradictions lead to no-go theorems for observer-independent
facts, thus challenging concepts of objectivity. Here, we examine
the paradoxes from the perspective of establishing consistency with
macroscopic realism. We present versions of the Brukner-Wigner-friend
and Frauchiger-Renner paradoxes in which the spin-$1/2$ system measured
by the Friends corresponds to two macroscopically distinct states.
The local unitary operations $U_{\theta}$ that determine the measurement
setting $\theta$ are carried out using nonlinear interactions, thereby
ensuring measurements need only distinguish between the macroscopically
distinct states. The macroscopic paradoxes are perplexing, seemingly
suggesting there is no objectivity in a macroscopic limit. However,
we demonstrate consistency with  a contextual weak form of macroscopic
realism (wMR): The premise wMR asserts that the system can be considered
to have a definite spin outcome $\lambda_{\theta}$, at the time after
the system has undergone the  unitary rotation $U_{\theta}$ to
prepare it in a suitable pointer basis. We further show that the paradoxical
outcomes imply failure of deterministic macroscopic local realism,
and arise when there are unitary interactions $U_{\theta}$ occurring
due to a change of measurement setting at both sites, with respect
to the state prepared by each Friend. In models which validate
wMR, there is a breakdown of a subset of the assumptions that constitute
the Bell-Locality premise. A similar interpretation involving a weak
contextual form of realism exists for the original paradoxes.
\end{abstract}
\maketitle

\section{Introduction}

The Wigner's friend paradox concerns a gedanken experiment in which
inconsistencies arise between observations recorded by experimentalists
either inside or outside the laboratory \citep{wigner-original}.
There is a distinction between systems that have undergone a ``collapse''
into an eigenstate due to measurement, and systems which remain entangled
with the laboratory apparatus. The inconsistencies can be quantified
in the form of Brukner's no-go theorem for ``observer-independent
facts'', that a record of the results of measurements exists in a
way that can be viewed consistently between observers \citep{brukner}.
The no-go theorem applies to an extended Wigner's friend paradox for
two laboratories, and adopts the Locality assumption. The inconsistencies
have been further highlighted by the Frauchiger-Renner paradox \citep{fr-paradox}.
As experiments support quantum predictions \citep{proietti,griffith},
the paradoxes challenge the concept of objectivity. This has motivated
much further work \citep{proietti,sudbery,bub,healey,bohmian-fr,griffith,losada-wigner-friend,fr-proposal-exp,wigner-weak,scalable,Lostaglio,Zukowski2021,Baumann2021,Castellani2021,Leegwater,Brukner2022}
including analyses involving consistent histories \citep{losada-wigner-friend},
Bohmian models \citep{bohmian-fr}, weak measurements \citep{wigner-weak},
timeless formulations \citep{Baumann2021}, and strong ``local friendliness''
no-go theorems \citep{griffith}.

In this paper, we present macroscopic versions of Brukner's Wigner's
friend and Frauchiger-Renner paradoxes in which \emph{all} measurements
leading to the inconsistent results are performed on a system ``which
has just two macroscopically distinguishable states available to it''
\citep{s-cat,frowis}. This includes the initial system measured by
the Friends in each laboratory, which is normally considered to be
microscopic. The consequence is that one can apply, for each system
and measurement, the definition of macroscopic realism put forward
by Leggett and Garg \citep{legggarg-1,emary-review}. Leggett and
Garg's macroscopic realism (MR) asserts that the system actually \emph{be}
in one or other state at any given time, meaning that the outcome
of a measurement distinguishing between the two macroscopically-distinct
states has a predetermined value.

With this definition, it would seem at first glance impossible to
obtain consistency between macroscopic realism and the Wigner's friend
paradoxes, which suggest there is no objectivity between observers
for the outcomes of quantum measurements on a macroscopic spin. The
paradoxes as applied to macroscopic qubits become especially puzzling,
because apparently then there is no basis for objectivity even in
a macroscopic limit.

In this paper, we examine the relationship of the paradoxes with MR,
giving a resolution of the apparent inconsistencies. We consider two
forms of MR, a deterministic form (dMR) which we show is negated by
the paradoxes, and a weaker form (wMR) which we show is consistent
with the quantum predictions, and is similar to Bell's idea of macroscopic
'beables' \citep{bell-found}. The resolution is based on the dynamics
of the unitary interaction $U_{\theta}$ that determines the measurement
settings for spin measurements $S_{\theta}$. In a contextual model
of quantum mechanics, the state after the interaction is different
to that before. We demonstrate that the premise of MR as applied to
the state \emph{after} the dynamics $U_{\theta}(t)$ takes place is
consistent with the quantum predictions: The premise of MR as applied
to the state \emph{before} the dynamics $U_{\theta}(t)$ takes place
is falsified by the quantum predictions. This leads to the two definitions
of MR \citep{manushan-bell-cat-lg,ghz-cat,delayed-choice-cats}. The
first is a weaker (more minimal) assumption, referred to as \emph{weak
macroscopic realism} (wMR). The second definition is a stronger assumption
that postulates predetermined variables prior to all stages of measurement,
along the lines of classical mechanics, and is referred to as \emph{deterministic
macroscopic realism} (dMR).

We therefore propose an interpretation that validates an objective
macroscopic realism, in which there is a predetermined value $\lambda_{\theta.i}$
for the outcome of the measurement on the macroscopic two-state system,
in accordance with wMR. The system is objectively in a state of definite
qubit value $\lambda_{\theta,i}$, at the time $t_{i}$ once the system
has undergone the appropriate unitary rotation $U_{\theta}$ to prepare
it in the suitable basis. The records of the Friends and the superobservers
agree on such values. The paradoxical outcomes between the two types
of observers (Friends and superobservers) illustrate a failure of
dMR, which we show manifests as a violation of a macroscopic Brukner-Wigner-Bell
inequality in both the extended Wigner's friend experiment, and the
Frauchiger-Renner version.

We further show that the inconsistencies between the different observers
arise where there are two nonzero unitary rotations, $U_{\theta}$
and $U_{\phi}$, giving a change of measurement setting $\theta$
and $\phi$ of the superobservers with respect to the Friends, at
\emph{both} available laboratories. In this case, the assumption of
Locality is justified by dMR, but \emph{not} by wMR. The predictions
that violate Brukner-Wigner-Bell or Bell inequalities are therefore
not inconsistent with wMR.

In fact, the premise of wMR implies a \emph{partial locality}, which
asserts no-disturbance to the value of $\lambda_{\theta,i}$ for the
state created at the time $t_{i}$\emph{, after} the local unitary
$U_{\theta}(t)$ has taken place. This is regardless of any unitary
interaction $U_{\phi}$ occurring at the other laboratory. However,
the premise wMR does not imply locality in the full sense: It cannot
be assumed that the outcome of a spin measurement $S_{\theta}$ at
one laboratory $A$ is independent of the measurement choice $\phi$
occurring at the other laboratory, if the local unitary $U_{\theta}$
at $A$ has not yet been performed.

The formulation of the macroscopic paradoxes is achieved by a direct
mapping of the microscopic gedanken experiment onto a macroscopic
one, the spin qubits $|$$\uparrow\rangle$ and $|\downarrow\rangle$
corresponding to two macroscopically distinct orthogonal states. We
illustrate with two examples: two coherent states $|\alpha\rangle$
and $|-\alpha\rangle$ where $\alpha\rightarrow\infty$, and two groups
of $N$ correlated spins. The unitary operations $U_{\theta}$ required
for the measurement of a spin component $S_{\theta}$ are realised
by a Kerr Hamiltonian $H_{NL}$, or else a sequence of CNOT gates.

The interpretation given in this paper  motivates a similar interpretation
for the original paradoxes, where the Friends make microscopic spin
measurements. In that interpretation, wMR is replaced by a weak version
of local realism (wR or wLR), which specifies a predetermined value
$\lambda_{\theta,i}$ for the outcome of a measurement $S_{\theta}$,
for the system prepared at time $t_{i}$ \emph{after} the unitary
interaction $U_{\theta}$ determining the choice of measurement setting
$\theta$ has been carried out. The interaction $U_{\theta}$ prepares
the system with respect to a measurement basis, in a state given as
the superposition $|\uparrow\rangle_{\theta}+|\downarrow\rangle_{\theta}$
of eigenstates of the spin observable $S_{\theta}$. In this contextual
model, the state is only completely described once the measurement
basis is specified. Similar contextual models have been given for
Bell violations \textcolor{black}{\citep{philippe-grang-context,manushan-bell-cat-lg}
and, in a full probabilistic formulation, for quantum measurement}
\citep{DrummondReid2020}.

\textbf{\emph{Overview of paper}}: The paper is organized as follows.
In Section II, we summarise the Wigner's friend and Frauchiger-Renner
gedanken paradoxes. We illustrate the fully macroscopic versions
of the paradoxes in Section III, where we show a violation of the
Brukner-Bell-Wigner inequality. In Section IV, we demonstrate the
failure of deterministic macroscopic (local) realism (dMR) for both
paradoxes. Weak macroscopic realism (wMR) is explained in Section
V, where it is shown how the weak form of realism can be compatible
with violations of the Brukner-Bell-Wigner and CHSH-Bell inequalities.
We prove a sequence of Results for wMR. In Section VI, the consistency
of the paradoxes with wMR is illustrated by way of an explicit wMR
model. This is done by comparing with the predictions of certain quantum
mixtures that are valid from one or other of the Friends' perspective.
A conclusion is given in Section VII.

\section{Wigner's friend paradoxes}

\subsection{Observer-independent facts no-go theorem: Bell-Wigner test}

\textcolor{red}{}We first summarise the theorem introduced by Brukner
for the Wigner's friend paradox \citep{brukner}. A spin-$1/2$ system
is in a closed laboratory $L$ where Wigner's friend $F$ can make
a measurement on a spin-$1/2$ system, to measure the $z$ component,
$\sigma_{z}$. This means that the spin system will become entangled
with a second more macroscopic system that exists in the laboratory.
(Ultimately, as each piece of apparatus becomes entangled, the macroscopic
apparatus becomes the Friend themselves). From the Friend's perspective,
after the measurement, the system has collapsed into a state that
has a definite value for the spin $S_{z}$ measurement. To Wigner,
who is outside the laboratory, the Friend's measurement is described
by a unitary interaction, where the combined state of the spin and
Friend is given by
\begin{align}
|\Phi\rangle_{SF} & =\frac{1}{\sqrt{2}}\left(|\uparrow\rangle_{z}|F_{z+}\rangle+|\downarrow\rangle_{z}|F_{z-}\rangle\right).\label{eq:wigner-friend}
\end{align}
Here, $|\uparrow\rangle_{z}$ and $|\downarrow\rangle_{z}$ are the
eigenstates of $\sigma_{z}$. The $|F_{z+}\rangle$ and $|F_{z-}\rangle$
are states for the macroscopic measuring apparatus, which we might
envisage to be a pointer on a measurement dial, that indicate the
result of the measurement to be either a positive outcome $+1$ (spin
``up''), or a negative outcome $-1$ (spin ``down'') respectively.
Wigner's description of the combined state is that of a superposition.
Hence, the interpretation of the overall state of the laboratory is
different, or unclear, since the superposition is not equivalent to
the mixture of the two states $|\uparrow\rangle_{z}|F_{z+}\rangle$
and $|\uparrow\rangle_{z}|F_{z+}\rangle$. 
\begin{figure}[t]
\begin{centering}
\includegraphics[width=1\columnwidth]{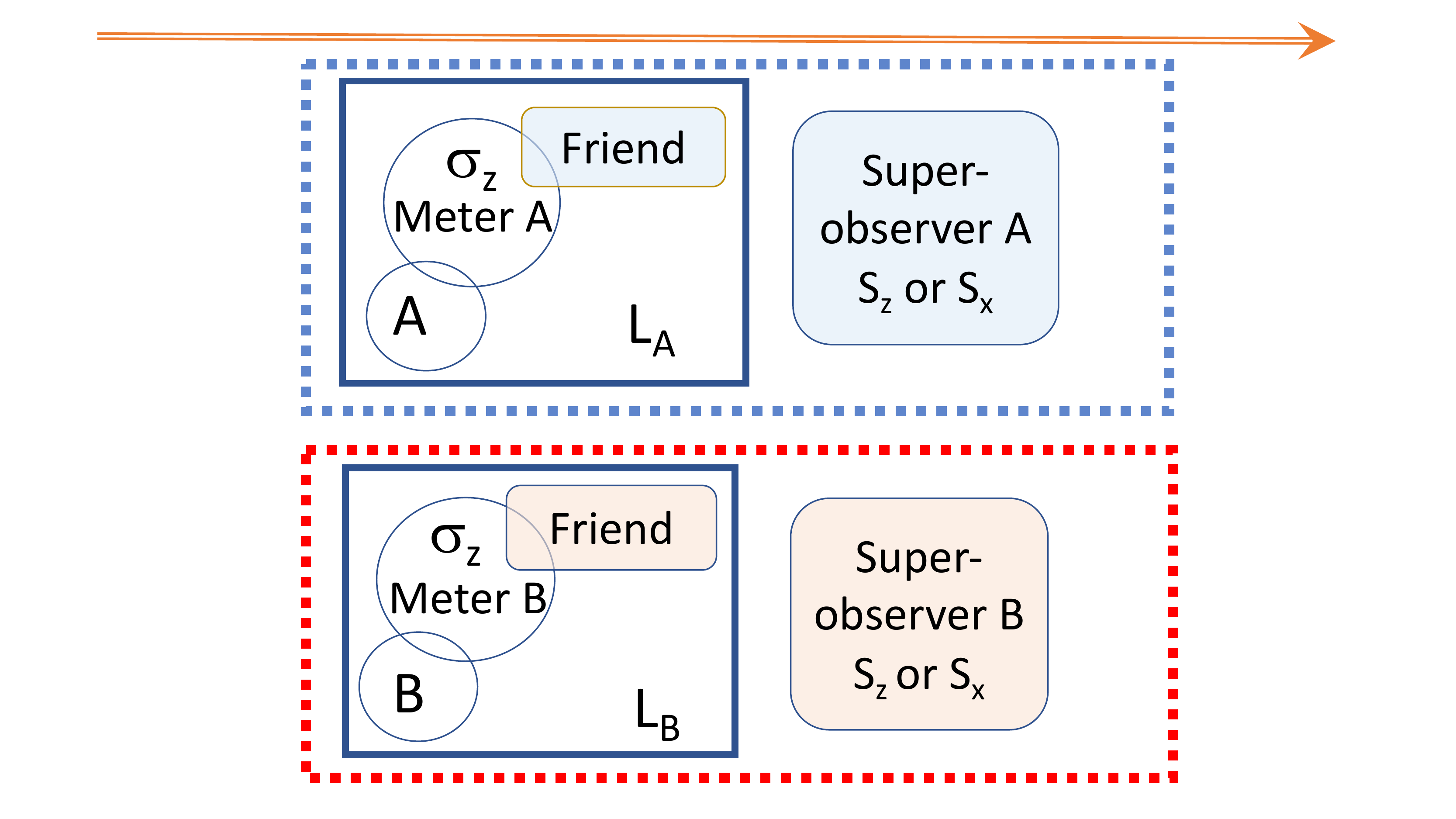}
\par\end{centering}
\caption{Wigner's friend paradox: The two entangled systems $A$ and $B$ are
prepared and then separated into laboratories $L_{A}$ and $L_{B}$.
The Friends in each laboratory measure the spin $\sigma_{z}$ of the
local system, using a macroscopic meter. The superobservers outside
each laboratory can measure the local macroscopic spins, $S_{z}$
or $S_{x}$, of the entire Lab systems. The arrow depicts the direction
of time, meaning that the superobservers make their measurements after
those of the Friends.\textcolor{blue}{\label{fig:Exp-diagram-1}}}
\end{figure}

A no-go theorem relating to the paradox was presented by Brukner,
who established a theoretical framework in which one can account for
observer-independent facts. The notion of observer-dependent facts
is tested by carrying out a Bell-Wigner experiment. Based on the work
of Brukner, a violation of a Bell-Wigner inequality implies a failure
in the conjunction of: (1) Locality; (2) Free choice (freedom for
all parties to choose their measurement settings; and (3) Observer-independent
facts (a record from a measurement should be a fact of the world that
all observers can agree on). The difference between a Bell test and
a Bell-Wigner test lies in the third assumption; a Bell-Wigner test
assumes observer-independent facts, while a Bell test assumes predetermined
measurement outcomes. Locality is defined by Bell in the original
derivation of Bell's inequalities, and implies no instantaneous influences
between spacelike-separated systems.

Brukner considered a pair of superobservers (Alice and Bob) w\textcolor{black}{ho
can carry out experiments on two separate laboratories $L_{A}$ and
$L_{B}$ that consist of spin-$1/2$ systems and the superobservers'
Friends, Charlie and Debbie, respectively (Figure \ref{fig:Exp-diagram-1}).
Measurement settings $A_{1}$ and $A_{2}$ correspond to the observational
statements of Charlie and Alice, while the measurement settings $B_{1}$
and $B_{2}$ correspond to the observational statements of Debbie
and Bob}. The conjunction of the assumptions \textcolor{black}{leads
to the Bell-Wigner inequality in the form of a Clauser-Horne-Shimony-Holt
(CHSH) Bell inequality \citep{cshim-review-2,bell-found,chsh,bell-brunner-rmp}
\begin{equation}
S=|\langle A_{1}B_{1}\rangle+\langle A_{1}B_{2}\rangle+\langle A_{2}B_{1}\rangle-\langle A_{2}B_{2}\rangle|\leq2.\label{eq:observer-independence}
\end{equation}
}A violation would imply a contradiction with the assumptions.\textcolor{black}{}

It has been shown that the inequality can be violated \citep{brukner}.
For our work, it will prove convenient to consider two strategies,
one involving measurements of $S_{z}$ and $S_{x}$ as in the original
example, and the other involving measurements of $S_{z}$ and $S_{y}$.
We therefore propose that Charlie and Debbie receive an entangled
state of spin-$1/2$ particles given as 
\begin{align}
|\psi_{\pm}\rangle & =-\sin\frac{\theta}{2}|\phi^{\mp}\rangle+\varepsilon_{\pm}\cos\frac{\theta}{2}|\psi^{+}\rangle\nonumber \\
 & =-\sin\frac{\theta}{2}{\color{black}(}{\color{black}|\uparrow\rangle_{z,C}|\uparrow\rangle_{z,D}\mp|\downarrow\rangle_{z,C}|\downarrow\rangle_{z,D})/\sqrt{2}}\nonumber \\
 & \ \ +\varepsilon_{\pm}\cos\frac{\theta}{2}{\color{black}(}{\color{black}|\uparrow\rangle_{z,C}|\downarrow\rangle_{z,D}+|\downarrow\rangle_{z,C}|\uparrow\rangle_{z,D})/\sqrt{2}}\label{eq:state}
\end{align}
where $|\phi^{\mp}\rangle=\frac{1}{\sqrt{2}}\left(|\uparrow\rangle_{z,C}|\uparrow\rangle_{z,D}\mp|\downarrow\rangle_{z,C}|\downarrow\rangle_{z,D}\right)$
and $|\psi^{+}\rangle=\frac{1}{\sqrt{2}}\left(|\uparrow\rangle_{z,C}|\downarrow\rangle_{z,D}+|\downarrow\rangle_{z,C}|\uparrow\rangle_{z,D}\right)$,
with $\varepsilon_{+}=1$ and $\varepsilon_{-}=i$. \textcolor{blue}{}The
subscripts $C$ and $D$ denote the spin states prepared in Charlie
and Debbie's laboratories. We define two initial states, $|\psi_{+}\rangle$
and $|\psi_{-}\rangle$, which will allow violation of the inequality
(\ref{eq:observer-independence}) for the pair of measurements $S_{z}$
and $S_{x}$, and the pair of measurements $S_{z}$ and $S_{y}$,
respectively.

Next, Charlie and Debbie each perform a measurement. After completing
their measurement on the system prepared in $|\psi_{\pm}\rangle$,
the overall state becomes
\begin{align}
|\tilde{\Psi}_{\pm}\rangle & =-\sin\frac{\theta}{2}|\Phi^{-}\rangle+\varepsilon_{\pm}\cos\frac{\theta}{2}|\Psi^{+}\rangle\,,\label{eq:state2}
\end{align}
where 
\begin{eqnarray}
|\Phi^{\mp}\rangle & = & \frac{1}{\sqrt{2}}\left(|A_{up}\rangle|B_{up}\rangle\mp|A_{down}\rangle|B_{down}\rangle\right)\nonumber \\
|\Psi^{+}\rangle & = & \frac{1}{\sqrt{2}}\left(|A_{up}\rangle|B_{down}\rangle+|A_{down}\rangle|B_{up}\rangle\right)\label{eq:state23}
\end{eqnarray}
with $|A_{up}\rangle=|\uparrow\rangle_{z,C}|C_{z+}\rangle_{C}$, $|A_{down}\rangle=|\downarrow\rangle_{z,C}|C_{z-}\rangle_{C}$,
$|B_{up}\rangle=|\uparrow\rangle_{z,D}|D_{z+}\rangle_{D}$, and $|B_{down}\rangle=|\downarrow\rangle_{z,D}|D_{z-}\rangle_{D}$.
Here, $|C_{z\pm}\rangle_{C}$ and $|D_{z\pm}\rangle_{D}$ are the
states of the macroscopic measurement apparatus (the Friends) in the
respective laboratories.

For the choice of initial state $|\psi_{+}\rangle$, we consider
the measurement settings 
\begin{eqnarray}
A_{1}\equiv A_{z} & = & |A_{up}\rangle\langle A_{up}|-|A_{down}\rangle\langle A_{down}|\nonumber \\
A_{2}\equiv A_{x} & = & |A_{up}\rangle\langle A_{down}|+|A_{down}\rangle\langle A_{up}|\label{eq:settingC}
\end{eqnarray}
corresponding to macroscopic spin $z$ and spin $x$ measurements
in Charlie's laboratory, and 
\begin{eqnarray}
B_{1}\equiv B_{z} & = & |B_{up}\rangle\langle B_{up}|-|B_{down}\rangle\langle B_{down}|\nonumber \\
B_{2}\equiv B_{x} & = & |B_{up}\rangle\langle B_{down}|+|B_{down}\rangle\langle B_{up}|\label{eq:settingD}
\end{eqnarray}
corresponding to macroscopic spin $z$ and spin $x$ measurements
in Debbie's laboratory. The Bell-Wigner CHSH inequality for this case
is\textcolor{black}{
\begin{equation}
S=|\langle A_{z}B_{z}\rangle+\langle A_{z}B_{x}\rangle+\langle A_{x}B_{z}\rangle-\langle A_{x}B_{x}\rangle|\leq2.\label{eq:observer-independence-2}
\end{equation}
}It has been shown that this inequality is violated for $\theta=\pi/4$
with $S=-2\sqrt{2}$.

On the other hand, for the choice of initial state $|\psi_{-}\rangle$,
we consider the measurement settings\textcolor{blue}{}
\begin{eqnarray}
A_{1}\equiv A_{z} & = & |A_{up}\rangle\langle A_{up}|-|A_{down}\rangle\langle A_{down}|\nonumber \\
A_{2}\equiv A_{y} & = & (|A_{up}\rangle\langle A_{down}|-|A_{down}\rangle\langle A_{up}|)/i\label{eq:settingC-1-1}
\end{eqnarray}
corresponding to macroscopic spin $z$ and spin $y$ measurements
in Charlie's laboratory, and 
\begin{eqnarray}
B_{1}\equiv B_{z} & = & |B_{up}\rangle\langle B_{up}|-|B_{down}\rangle\langle B_{down}|\nonumber \\
B_{2}\equiv B_{y} & = & (|B_{up}\rangle\langle B_{down}|-|B_{down}\rangle\langle B_{up}|)/i\label{eq:settingD-1-1}
\end{eqnarray}
corresponding to macroscopic spin $z$ and spin $y$ measurements
in Debbie's laboratory. The Bell-Wigner CHSH inequality for this case
i\textcolor{black}{s
\begin{equation}
S=\langle A_{z}B_{z}\rangle+\langle A_{z}B_{y}\rangle+\langle A_{y}B_{z}\rangle-\langle A_{y}B_{y}\rangle\leq2\label{eq:observer-independence-2-1-1}
\end{equation}
}This the case of interest in this paper. We evaluate the correlations
as follows. \textcolor{blue}{}Directly, we find
\[
\langle A_{z}B_{z}\rangle=-\cos\theta.
\]
To evaluate $\langle A_{z}B_{y}\rangle$, we write the state in the
new basis, by noting the standard transformation
\begin{eqnarray}
|\uparrow\rangle_{y} & = & \frac{1}{\sqrt{2}}(|\uparrow\rangle_{z}+i|\downarrow\rangle_{z})\nonumber \\
|\downarrow\rangle_{y} & = & \frac{1}{\sqrt{2}}(|\uparrow\rangle_{z}-i|\downarrow\rangle_{z})\label{eq:transy-1-1-1}
\end{eqnarray}
where $|\uparrow\rangle_{y}$ and $|\downarrow\rangle_{y}$ are the
eigenstates of the Pauli spin $\sigma_{y}$, with eigenvalues $+1$
and $-1$ respectively. Hence $|\uparrow\rangle_{z}=(|\uparrow\rangle_{y}+|\downarrow\rangle_{y})/\sqrt{2}$
and $|\downarrow\rangle_{z}=-i(|\uparrow\rangle_{y}-|\downarrow\rangle_{y})/\sqrt{2}$.
Substituting, we find
\begin{eqnarray*}
\langle A_{z}B_{y}\rangle & = & \langle A_{y}B_{z}\rangle=-\sin\theta\\
\langle A_{y}B_{y}\rangle & = & \cos\theta.
\end{eqnarray*}
\textcolor{blue}{}The Bell-Wigner inequality is violated with
$|S|=2\sqrt{2}$ for $\theta=\pi/4$. An experimental test supporting
the predictions of quantum mechanics has been carried out by Proietti
et al. \citep{proietti}.\textcolor{red}{}\textcolor{blue}{}

\subsection{Frauchiger-Renner paradox}

Here we outline the Frauchiger-Renner paradox \citep{fr-paradox}
which also examines the Wigner friend's thought experiment, arriving
at a contradiction between the Friends inside the laboratories and
the observers outside.  We follow the summary given by Losada et
al. \citep{losada-wigner-friend}.

First, a biased quantum coin tossed by the Friend $F_{A}$ in laboratory
$A$ gives outcomes $h$ and $t$ with probabilities $1/3$ and $2/3$
respectively. If the outcome is $h$ or $t$, the Friend $F_{B}$
in the second laboratory $L_{B}$ creates the spin $1/2$ state $|\downarrow\rangle$
or $|\rightarrow\rangle=\frac{1}{\sqrt{2}}(|\uparrow\rangle+|\downarrow\rangle)$
respectively. Here, $|h\rangle_{z}$, $|t\rangle_{z}$ and $|\uparrow\rangle_{z}$,
$|\downarrow\rangle_{z}$ are the eigenstates of the Pauli spin observables
$\sigma_{z}^{A}$ and $\sigma_{z}^{B}$ for two spin $1/2$ systems
at the spatially separated laboratories $L_{A}$ and $L_{B}$, respectively.

In fact, the friend $F_{A}$ has measured the state of the coin, by
first coupling with a device in $L_{A}$, later measured by the Friend.
The macroscopic state ultimately represents all macroscopic devices
leading to the measurement outcome. The $|H\rangle_{z}$ or $|T\rangle_{z}$
are eigenstates associated with the values $h$ and $t$, for the
overall laboratory $L_{A}$. The $|H\rangle_{z}$ and $|T\rangle_{z}$
are eigenstates of the observable denoted $S_{z}^{A}$. The coupling
to the second laboratory $L_{B}$ is described by an interaction Hamiltonian.
The system is coupled to a system in the second laboratory $B$, so
that a final overall entangled state\textcolor{blue}{}
\begin{equation}
|\psi_{zz}\rangle_{FR}=\frac{1}{\sqrt{3}}|H\rangle_{z}|\Downarrow\rangle_{z}+\sqrt{\frac{2}{3}}|T\rangle_{z}|\Rightarrow\rangle_{z}\label{eq:we}
\end{equation}
is created. Here, $|\Rightarrow\rangle_{z}=\frac{1}{\sqrt{2}}(|\Uparrow\rangle_{z}+|\Downarrow\rangle_{z})$,
where $|\Uparrow\rangle_{z}$ and $|\Downarrow\rangle_{z}$ are the
overall eigenstates of the $L_{B}$ associated with the final outcomes
of $\sigma_{z}^{B}$. The $|\Uparrow\rangle_{z}$ and $|\Downarrow\rangle_{z}$
are eigenstates of the observable denoted $S_{z}^{B}$.

The second step is that the external superobservers $W_{A}$ and $W_{B}$
make measurements of $S_{x}$, on the systems in $L_{A}$ and $L_{B}$
respectively. These observables are defined, so that the eigenstates
of $S_{x}^{A}$ are for laboratory $L_{A}$,
\begin{eqnarray}
|H\rangle_{x} & = & \frac{1}{\sqrt{2}}|H\rangle_{z}+\frac{1}{\sqrt{2}}|T\rangle_{z}\label{eq:failxokx}\\
|T\rangle_{x} & = & \frac{1}{\sqrt{2}}|H\rangle_{z}-\frac{1}{\sqrt{2}}|T\rangle_{z},\nonumber 
\end{eqnarray}
and for $S_{x}^{B}$ of laboratory $L_{B}$,
\begin{eqnarray}
|\Uparrow\rangle_{x} & = & \frac{1}{\sqrt{2}}|\Uparrow\rangle_{z}+\frac{1}{\sqrt{2}}|\Downarrow\rangle_{z}\label{eq:failyoky}\\
|\Downarrow\rangle_{x} & = & \frac{1}{\sqrt{2}}|\Uparrow\rangle_{z}-\frac{1}{\sqrt{2}}|\Downarrow\rangle_{z}.\nonumber 
\end{eqnarray}
We first consider where $W_{A}$ and $W_{B}$ would both measure $S_{x}$.
We rewrite the state (\ref{eq:we}) in terms of the different bases.
We have in the new basis
\begin{eqnarray}
|\psi_{xx}\rangle & = & \frac{\sqrt{3}}{2}|H\rangle_{x}|\Uparrow\rangle_{x}-\frac{1}{\sqrt{12}}|T\rangle_{x}|\Uparrow\rangle_{x}\nonumber \\
 &  & \ \ \ \ -\frac{1}{\sqrt{12}}|H\rangle_{x}|\Downarrow\rangle_{x}-\frac{1}{\sqrt{12}}|T\rangle_{x}|\Downarrow\rangle_{x}.\label{eq:xx}
\end{eqnarray}
\textcolor{blue}{}From the above expression, we see that the probability
is $\frac{1}{12}$ for obtaining outcomes $T$ and $\Downarrow$.

But now, if the friend $F_{B}$ had measured $\sigma_{z}$, and $W_{A}$
measures $S_{x}^{A}$ respectively, we would write
\begin{equation}
|\psi_{xz}\rangle=\frac{1}{\sqrt{6}}|H\rangle_{x}|\Uparrow\rangle_{z}-\frac{1}{\sqrt{6}}|T\rangle_{x}|\Uparrow\rangle_{z}+\frac{2}{\sqrt{6}}|H\rangle_{x}|\Downarrow\rangle_{z}.\label{eq:XZold}
\end{equation}
Here, the probability to get $|T\rangle_{x}|\Downarrow\rangle_{z}$
is zero. This implies, when $W_{A}$ obtains $T$ for $S_{x}$, they
can confirm with certainty that $F_{B}$ would have obtained $\uparrow$
for the measurement of $\sigma_{z}$, which in turn would seemingly
imply for friend $F_{B}$ that $F_{A}$ \emph{had} obtained $T$ for
$\sigma_{z}$ i.e. $\lambda_{z}^{A}=-1$ (since the $\uparrow$ state
for $\sigma_{z}$ at system $B$ is only created when $F_{A}$ would
have obtained $T$ for $\sigma_{z}$). \textcolor{teal}{} In the
basis of $\sigma_{z}$ for $L_{A}$ and $S_{x}$ for $L_{B}$, the
state is
\begin{equation}
|\psi_{zx}\rangle=\frac{1}{\sqrt{6}}|H\rangle_{z}\left(|\Uparrow\rangle_{x}-|\Downarrow\rangle_{x}\right)+\sqrt{\frac{2}{3}}|T\rangle_{z}|\Uparrow\rangle_{x}.\label{eq:ZXold-1}
\end{equation}
As the outcome $T$ for $\sigma_{z}$ is perfectly correlated with
the state $|\Uparrow\rangle_{x}$, it is certain from (\ref{eq:ZXold-1})
that $W_{B}$ would then get $\Uparrow$ for their measurement $S_{x}$
in lab $L_{B}.$ But (\ref{eq:xx}) gives a nonzero probability for
$W_{A}$ and $W_{B}$ getting $T$ and $\Downarrow$ for $S_{x}$,
and this is the basis of FR paradox.

\section{Wigner's friend paradoxes using cat states}

\begin{figure}[t]
\begin{centering}
\includegraphics[width=1\columnwidth]{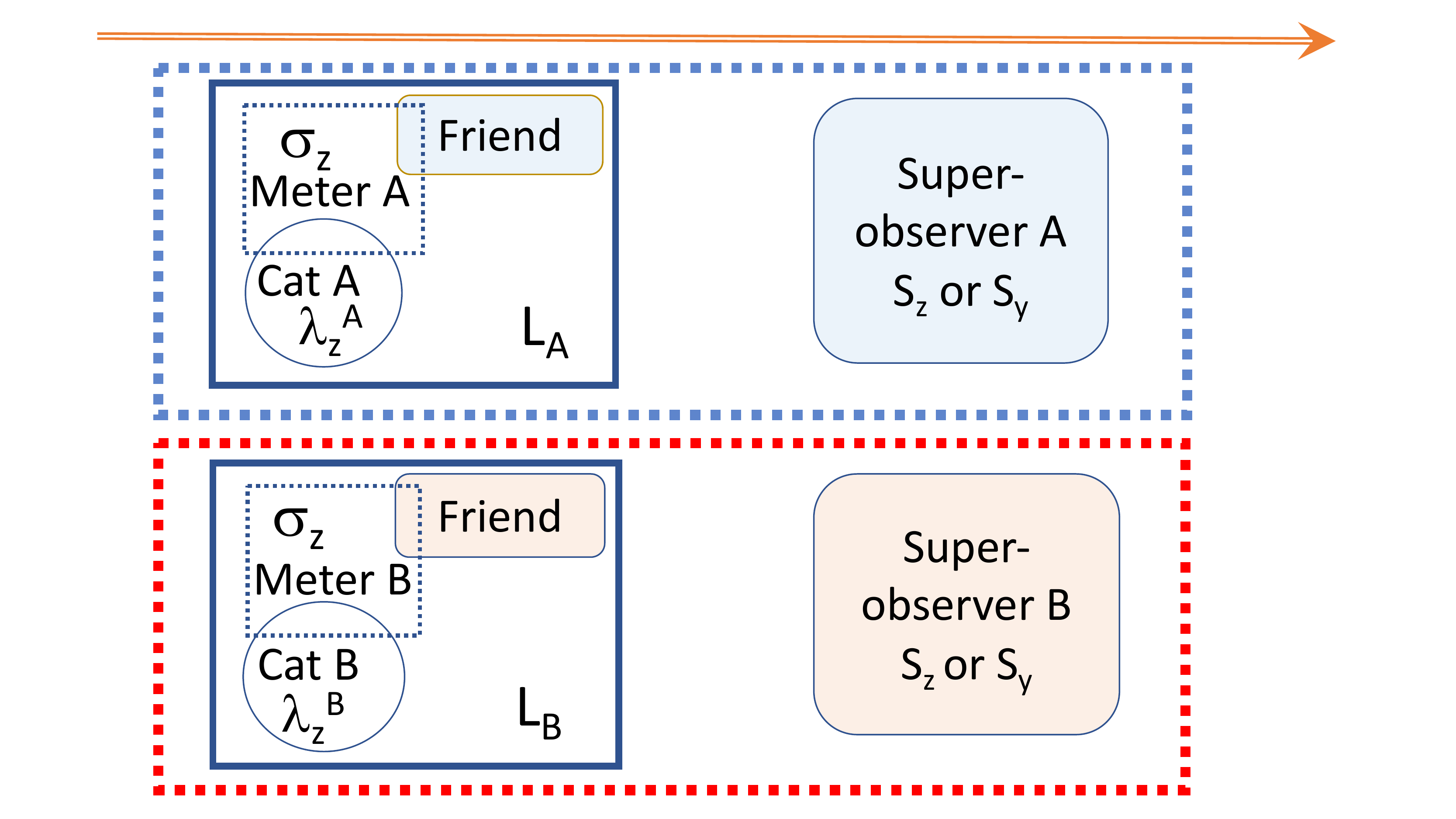}
\par\end{centering}
\caption{A macroscopic paradox with cat states: The two entangled systems $A$
and $B$ are prepared and then separated into laboratories $L_{A}$
and $L_{B}$, as in Figure 1. Here, the systems $A$ and $B$ are
themselves macroscopic, meaning that the spins values $+1$ and $-1$
for $\sigma_{z}$ correspond to macroscopically distinct states. The
premise of macroscopic realism asserts that the value of the spin
outcome $\sigma_{z}$ is predetermined, given by a variable $\lambda_{z}$.
\textcolor{blue}{\label{fig:Exp-diagram-cat}}}
\end{figure}

In this section, we present the paradox in terms of cat states \citep{frowis,s-cat},
where the \emph{original} spin $1/2$ system measured by the Friends
is a macroscopic one, and the two spin $1/2$ eigenstates are macroscopically
distinct states (macroscopic qubits). This is depicted in Figure \ref{fig:Exp-diagram-cat}.

An important aspect of the Wigner's friend paradox is that the measurement
of spin occurs in two stages, one of which is reversible. In the cat
example, this is also true, and we will be examining a specific macroscopic
realisation, so that we are able to analyse the dynamics of the measurements.

The first stage of the spin measurement is the unitary interaction
$U_{\theta}$ which determines the measurement setting i.e. the component
of spin that will be measured, whether $\sigma_{z}$, $\sigma_{x}$
or $\sigma_{y}$. This stage transforms an initial eigenstate into
a superposition of two eigenstates: e.g. $|\uparrow\rangle\rightarrow\frac{1}{\sqrt{2}}(|\uparrow\rangle+|\downarrow\rangle)$.
In a photonic Bell experiment, the unitary transformations are achieved
by polarising beam splitters (PBS). The unitary rotations $U_{\theta}$
for coherent-state qubits are explained below. This stage of measurement
is reversible.

The second stage of measurement occurs after the unitary rotation
$U_{\theta}$. Once the unitary rotation has been performed, the measurement
setting has been selected, and the system is prepared for a final
stage of measurement, that we will refer to as the ``pointer'' measurement.
This would usually include a final amplification and detection stage,
involving a coupling to a meter, and a read-out to a second system
e.g. an observer. We might also refer to this stage of measurement
as the ``collapse'' stage, because, from the perspective of the
observer making the measurement, this stage is irreversible.\textcolor{red}{}

The paradoxes involve spin measurements on each system, $A$ and $B$.
For macroscopic qubits, the spin measurements are defined according
to the two-level operators. For example, the macroscopic two-state
systems of the entire Labs are denoted by $|\Uparrow\rangle$ and
$|\Downarrow\rangle$, and $|H\rangle$ and $|T\rangle$. The corresponding
two-state spin observables are
\begin{eqnarray}
S_{z}^{A} & = & |H\rangle\langle H|-|T\rangle\langle T|\nonumber \\
S_{x}^{A} & = & |H\rangle\langle T|-|T\rangle\langle H|\nonumber \\
S_{y}^{A} & = & \{|H\rangle\langle T|-|H\rangle\langle T|\}/i\label{eq:spin1}
\end{eqnarray}
and
\begin{eqnarray}
S_{z}^{B} & = & |\Uparrow\rangle\langle\Uparrow|-|\Downarrow\rangle\langle\Downarrow|\nonumber \\
S_{x}^{B} & = & |\Uparrow\rangle\langle\Downarrow|-|\Downarrow\rangle\langle\Uparrow|\nonumber \\
S_{y}^{B} & = & \{|\Uparrow\rangle\langle\Downarrow|-|\Uparrow\rangle\langle\Downarrow|\}/i.\label{eq:spin2}
\end{eqnarray}
The paradoxes concern noncommuting spin measurements. We therefore
will seek a unitary transformation $U_{x}^{-1}$ that transforms the
eigenstates of $S_{z}$ into eigenstates of $S_{x}$:
\begin{eqnarray}
|H\rangle & \rightarrow & (|H\rangle+|T\rangle)/\sqrt{2}\nonumber \\
|T\rangle & \rightarrow & (|H\rangle-|T\rangle)/\sqrt{2}\label{eq:ransx}
\end{eqnarray}
or else the transformation $U_{y}^{-1}$ that transforms the eigenstates
of $S_{z}$ into eigenstates of $S_{y}$:
\begin{eqnarray}
|H\rangle & \rightarrow & (|H\rangle+i|T\rangle)/\sqrt{2}\nonumber \\
|T\rangle & \rightarrow & (|H\rangle-i|T\rangle)/\sqrt{2}\label{eq:transy}
\end{eqnarray}
(and similarly for $|\Uparrow\rangle$ and $|\Downarrow\rangle$).

For the initial spin $1/2$ system measured by the Friends, we propose
three sorts of macroscopic qubit. The first two are presented in Appendix
A. For the third, we consider the spins $|\uparrow\rangle$ and $|\downarrow\rangle$
to be macroscopic coherent states $|\alpha\rangle$ and $|-\alpha\rangle$,
where $\alpha$ is large and real (Figure \ref{fig:Exp-diagram-cat}).
In the limit $\alpha\rightarrow\infty$, the two states are orthogonal,
and one defines two-state spin observables, for Lab $A$, as:
\begin{eqnarray}
\sigma_{z}^{A} & = & |\alpha\rangle\langle\alpha|-|-\alpha\rangle\langle-\alpha|\nonumber \\
\sigma_{x}^{A} & = & |\alpha\rangle\langle-\alpha|-|-\alpha\rangle\langle\alpha|\nonumber \\
\sigma_{y}^{A} & = & \{|\alpha\rangle\langle-\alpha|-|\alpha\rangle\langle-\alpha|\}/i.\label{eq:spin2-1}
\end{eqnarray}
There is a direct mapping between the qubits $|\uparrow\rangle$
and $|\downarrow\rangle$ and the macroscopic qubits $|\alpha\rangle$
and $|-\alpha\rangle$. Similarly,
\begin{eqnarray}
\sigma_{z}^{B} & = & |\beta\rangle\langle\beta|-|-\beta\rangle\langle-\beta|\nonumber \\
\sigma_{x}^{B} & = & |\beta\rangle\langle-\beta|-|-\beta\rangle\langle\beta|\nonumber \\
\sigma_{y}^{B} & = & \{|\beta\rangle\langle-\beta|-|\beta\rangle\langle-\beta|\}/i\label{eq:spin2-1-1}
\end{eqnarray}
where $|\beta\rangle$ and $|-\beta\rangle$ ($\beta\rightarrow\infty$)
are macroscopically distinct coherent states for a mode in Lab $B$.
 We consider quadrature phase amplitude observables
\begin{eqnarray}
\hat{X}_{A} & = & (\hat{a}+\hat{a}^{\dagger})/2\nonumber \\
\hat{P}_{A} & = & (\hat{a}-\hat{a}^{\dagger})/2i\label{eq:quad-1}
\end{eqnarray}
defined for a single field mode in a rotating frame, where  $\hat{a}$
is the destruction operator for a system $A$ \citep{yurke-stoler-1}.
The two states $|\alpha\rangle$ and $|-\alpha\rangle$ can be distinguished
by a measurement of $\hat{X}_{A}$. The sign of the outcome gives
the qubit value, whether $+1$ or $-1$. The measurement $\hat{X}_{A}$
constitutes the pointer measurement of the system. Similar observables
$\hat{X}_{B}$ and $\hat{P}_{B}$ are defined for a mode $B$.

It will be necessary to also consider how to realise the first part
of the measurement process, which determines whether $S_{z}$, $S_{y}$
or $S_{x}$ will be measured by the observers. The unitary transformation
for $U_{y}^{A}$ can be achieved using a Kerr nonlinearity, modelled
by the Hamiltonian
\begin{equation}
H_{NL}^{A}=\hbar\Omega\hat{n}_{A}^{2}\label{eq:mint}
\end{equation}
where $\hat{n}_{A}=\hat{a}^{\dagger}\hat{a}$ is the number operator.
After an interaction time $t=\pi/2\Omega$, the system initially prepared
in a coherent state $|\alpha\rangle$ becomes a cat state. We find
\citep{yurke-stoler-1}
\begin{eqnarray}
U_{A}(\frac{\pi}{2\Omega})|\alpha\rangle & = & \frac{e^{-i\pi/4}}{\sqrt{2}}(|\alpha\rangle+i|-\alpha\rangle)\label{eq:cat}
\end{eqnarray}
where $U_{A}(t)=e^{-iH_{NL}^{A}t/\hbar}$. We use the notation $U_{A}\equiv U_{\pi/4}^{A}\equiv U_{A}(\frac{\pi}{2\Omega})$
to denote the transformation. Hence, for (\ref{eq:transy}), we select
$U_{y}=U_{A}^{-1}=(U_{\pi/4}^{A})^{-1}$. The cat states have been
created for a microwave field, using a dispersive Kerr interaction
$H_{NL}^{A}$ \citep{collapse-revival-super-circuit-1,cat-states-super-cond},
and similar effects are observed in Bose-Einstein condensates \citep{collapse-revival-bec-2,wrigth-walls-gar-1}.
Hence, the unitary interactions associated with the measurement $S_{y}$
are performed via the inverse of $U_{\pi/4}^{A}$ and $U_{\pi/4}^{B}$.
Thus, we write
\begin{eqnarray}
|\pm\rangle_{y,A} & =U_{\pi/4}^{A}|\pm\alpha\rangle_{z}= & \frac{e^{-i\pi/4}}{\sqrt{2}}(|\pm\alpha\rangle_{z}+i|\mp\alpha\rangle_{z})\nonumber \\
|\pm\rangle_{y,B} & =U_{\pi/4}^{B}|\pm\beta\rangle_{z}= & \frac{e^{-i\pi/4}}{\sqrt{2}}(|\pm\beta\rangle_{z}+i|\mp\beta\rangle_{z})\nonumber \\
\label{eq:eigenstate-y-1-1-1}
\end{eqnarray}
where we use that $U_{A}(t)=e^{-iH_{NL}^{A}t/\hbar}$ and $U_{B}(t)=e^{-iH_{NL}^{B}t/\hbar}$
as in (\ref{eq:cat}). The different overall phase compared to the
definition (\ref{eq:transy}) does not change that the states are
eigenstates of $\sigma_{y}$. This gives the required transformation,
 on denoting $|\pm\rangle_{z,A}\equiv|\pm\alpha\rangle_{z}$ and
$|\pm\rangle_{y,A}\equiv|\pm\alpha\rangle_{y,A}$. It is straightforward
to verify that $|\pm\rangle_{y,A}$ are the eigenstates of $\sigma_{y}^{A}$,
given by Eq. (\ref{eq:spin2-1}) where $\alpha\rightarrow\infty$.

To rewrite the basis states for $z$ in the basis for $y$, we operate
on the states by $U_{A}^{-1}$ ($U_{B}^{-1}$):
\begin{eqnarray}
|\pm\alpha\rangle_{z} & = & U_{A}^{-1}|\pm\alpha\rangle_{y}\nonumber \\
 & = & \frac{e^{i\pi/4}}{\sqrt{2}}(|\pm\alpha\rangle_{y}-i|\mp\alpha\rangle_{y}).\label{eq:eigenstate-y-1-1-2-2}
\end{eqnarray}
More generally, for a state $|\psi\rangle$ written as a superposition
of the eigenstates of $\sigma_{z}$, the transformation into the eigenstates
of $\sigma_{y}$ is given by
\begin{equation}
U_{y}|\psi\rangle\equiv U_{A}^{-1}|\psi\rangle.\label{eq:trans}
\end{equation}
A similar local transformation takes place on system $B$.

The odd and even cat states $\sim|\alpha\rangle\pm|-\alpha\rangle$
which would correspond to the transformation for $U_{x}$ have also
been created in the laboratory \citep{odd-cats}, but proposed mechanisms
for generation involve conditional measurements and dissipative optical,
superconducting or opto-mechanical systems \citep{transient-cat-states-leo,cat-even-odd-transient,cats-hach,cat-states-wc,Teh2018Creation,Teh2020Dynamics}.
In this paper, we focus on the cat states generated by the simple
unitary transformation $U_{y}^{A}$ that is realisable using $H_{NL}^{A}$.
This means we focus on the versions of paradoxes that use $S_{y}$
rather than $S_{x}$ measurements. The coherent-state qubits and unitary
interactions $U_{y}$ also allow macroscopic Bell violations \citep{manushan-bell-cat-lg,macro-bell-lg},
tests of macrorealism \citep{manushan-cat-lg}, macroscopic GHZ paradoxes
\citep{ghz-cat}, and tests of two-dimensional macroscopic retrocausal
models in delayed-choice Wheeler-Chaves-Lemos-Pienaar experiments
\citep{delayed-choice-cats}.

\subsection{Bell-Wigner tests with cat states}

We now propose that the Bell-Wigner test given in Section II.A be
implemented with the spins $|\uparrow\rangle_{z,C}$ and $|\downarrow\rangle_{z,C}$
realised as the macroscopic coherent states $|\alpha\rangle_{z,c}$
and $|-\alpha\rangle_{z,C}$, and the spins $|\uparrow\rangle_{z,D}$
and $|\downarrow\rangle_{z,D}$ realised as the macroscopic states
$|\alpha\rangle_{z,D}$ and $|-\alpha\rangle_{z,D}$, for large $\alpha$.
We explicitly write the state $|\psi_{-}\rangle$ given by Eq. (\ref{eq:state})
as
\begin{align}
|\psi_{zz}\rangle_{WF} & =A_{zz}{\color{black}(}{\color{black}|\alpha\rangle_{z,C}|\alpha\rangle_{z,D}+|-\alpha\rangle_{z,C}|-\alpha\rangle_{z,D})}\nonumber \\
 & \ \ +{\color{black}B_{zz}(|\alpha\rangle_{z,C}|-\alpha\rangle_{z,D}+|-\alpha\rangle_{z,C}|\alpha\rangle_{z,D})}\label{eq:state-1-2}
\end{align}
where $A_{zz}=-\frac{1}{\sqrt{2}}\sin\frac{\theta}{2}$ and $B_{zz}=\frac{i}{\sqrt{2}}\cos\frac{\theta}{2}$.
This state describes the system prepared in the $z$ basis at both
sites (Labs). A method for mapping the state (\ref{eq:state}) onto
the coherent-state version (\ref{eq:state-1-2}) experimentally is
presented in Refs. \citep{cat-det-map,cat-bell-wang-1}.

Charlie and Debbie then each perform a measurement. This involves
coupling each system with a meter via interactions $H_{Am}$ and $H_{Bm}$,
and then further couplings to the Friends in each laboratory (Lab).
The overall interaction of systems with the macroscopic apparatus
are described by $H_{AmF}$ and $H_{BmF}$. After completing their
measurements, the overall state of the Labs becomes $|\tilde{\Psi}_{-}\rangle=-\sin\frac{\theta}{2}|\Phi^{+}\rangle+i\cos\frac{\theta}{2}|\Psi^{+}\rangle\,,$where
$|\Phi^{+}\rangle$ and $|\Psi^{+}\rangle$ are given by Eq. (\ref{eq:state23}),
with $|A_{up}\rangle=|\alpha\rangle_{z,C}|C_{z+}\rangle_{C}$, $|A_{down}\rangle=|-\alpha\rangle_{z,C}|C_{z-}\rangle_{C}$,
$|B_{up}\rangle=|\alpha\rangle_{z,D}|D_{z+}\rangle_{D}$, and $|B_{down}\rangle=|-\alpha\rangle_{z,D}|D_{z-}\rangle_{D}$.
Here, $|C_{z\pm}\rangle_{C}$ and $|D_{z\pm}\rangle_{D}$ are the
states of the macroscopic measurement apparatus (the friends) in the
respective Labs. An example of the measurement interaction $H_{Am}$
is given in the Appendix.

At each Lab, the superobservers have the choice to measure either
$S_{z}$ or $S_{y}$. To measure $S_{y}$, they first disentangle
the system from the respective meters (by reversing $H_{AmF}$ or
$H_{BmF}$), and then perform the unitary transformations $U_{A}^{-1}$
and $U_{B}^{-1}$ as in (\ref{eq:eigenstate-y-1-1-2-2}). A pointer
measurement is then needed to give the final readout for $S_{y}$,
meaning that the evolved system is coupled once more to the measurement
apparatus, in the superobserver's Lab.

To measure $S_{z}$, no further unitary rotation $U$ is needed, because
the system given by (\ref{eq:state-1-2}) is already prepared in the
basis for spin $Z$. A pointer measurement suffices to determine the
final outcome for $S_{z}$. However, for simplicity of treatment,
 it is convenient to consider that the superobservers will in any
case reverse both the couplings $H_{AmF}$ and $H_{BmF}$, regardless
of their choice of measurement. This decouples the spin system from
the Lab measurement apparatus (the meter and Friends), and allows
a simple description of the spin systems as they evolve under any
unitary rotations. In this case, a pointer measurement at a later
time couples the evolved system to the meters, and measurement apparatus
in the superobserver's Lab. The final result for $S_{z}$ will agree
with that of the Friend.

Consider measurements of $S_{y}^{A}$ and $S_{z}^{B}$. After reversing
$H_{AmF}$ and $H_{BmF}$ and performing the necessary unitary rotations,
the system is given by
\begin{align}
|\psi_{zy}\rangle_{WF} & =A_{zy}\{|\alpha\rangle_{z,C}|\alpha\rangle_{y,D}+i|-\alpha\rangle_{z,C}|-\alpha\rangle_{y,D}\}\nonumber \\
 & \ \ +B_{zy}(-|\alpha\rangle_{z,C}|-\alpha\rangle_{y,D}+i|-\alpha\rangle_{z,C}|\alpha\rangle_{y,D})\label{eq:state-1-2-1}
\end{align}
where $A_{zy}=\frac{1}{2}(-\sin\frac{\theta}{2}+\cos\frac{\theta}{2})$
and $B_{zy}=\frac{1}{2}(\sin\frac{\theta}{2}+\cos\frac{\theta}{2})$.
Similarly, the state of the system prepared for measurements $S_{z}^{A}$
and $S_{y}^{B}$ is
\begin{align}
|\psi_{yz}\rangle_{WF} & =A_{yz}\{|\alpha\rangle_{y,C}|\alpha\rangle_{z,D}+i|-\alpha\rangle_{y,C}|-\alpha\rangle_{z,D}\}\nonumber \\
 & \ \ +B_{yz}(i|\alpha\rangle_{y,C}|-\alpha\rangle_{z,D}-|-\alpha\rangle_{y,C}|\alpha\rangle_{z,D})\label{eq:state-1-2-1-1}
\end{align}
where $A_{yz}=\frac{1}{2}(-\sin\frac{\theta}{2}+\cos\frac{\theta}{2})$
and $B_{yz}=\frac{1}{2}(\sin\frac{\theta}{2}+\cos\frac{\theta}{2})$.
Similarly, for the measurements $S_{y}^{A}$ and $S_{y}^{B}$,
\begin{align}
|\psi_{yy}\rangle_{WF} & =A_{yy}{\color{black}(}{\color{black}|\alpha\rangle_{y,C}|\alpha\rangle_{y,D}-|-\alpha\rangle_{y}|-\alpha\rangle_{y,C})}\nonumber \\
 & \ \ +B_{yy}{\color{black}(}{\color{black}|\alpha\rangle_{y,C}|-\alpha\rangle_{y,D})+|-\alpha\rangle_{y,C}|\alpha\rangle_{y,D})}\label{eq:state-1-2-1-1-1}
\end{align}
where $B_{yy}=-\frac{1}{\sqrt{2}}\sin\frac{\theta}{2}$ and $A_{yy}=\frac{1}{\sqrt{2}}\cos\frac{\theta}{2}$.
Comparing with Section II.A, this gives a violation of the Bell-Wigner
inequality.

\subsection{Frauchiger-Renner Paradox using cat states}

We now consider a test of the FR paradox based on cat states. The
version of the FR paradox given in Section II.C considers the option
that the observers measure either $S_{z}$ or $S_{x}$. We suppose
instead that the observers at $A$ and $B$ measure either $S_{z}$
or $S_{y}$, so that we may use the transformation (\ref{eq:cat}).
To do this, we consider that the state initially prepared between
the laboratories is
\begin{eqnarray}
|\psi_{zz}\rangle & = & \frac{1}{\sqrt{3}}|H\rangle_{z,A}|\Downarrow\rangle_{z,B}+\frac{1}{\sqrt{3}}|T\rangle_{A}(|\Uparrow\rangle_{z,B}+i|\Downarrow\rangle_{z,B}).\nonumber \\
\label{eq:fr1}
\end{eqnarray}
We suppose the state is prepared with respect to the basis $S_{z}$
at each location, $A$ and $B$. A paradox can be constructed the
same way as explained in Section II.B above, where the measurements
of $S_{x}$ are substituted as measurements of $S_{y}$.

We now present a version of the paradox using coherent states. We
suppose the initial state created between the two laboratories is
\begin{eqnarray}
|\psi_{zz}\rangle_{FR} & = & \frac{1}{\sqrt{3}}|\alpha\rangle_{z}|-\beta\rangle_{z}\nonumber \\
 &  & \ \ \ +\frac{1}{\sqrt{3}}|-\alpha\rangle_{z}(|\beta\rangle_{z}+i|-\beta\rangle_{z}).\label{eq:ini}
\end{eqnarray}
We drop the subscripts $A$ and $B$, where it is clear that the first
(second) ket refers to the first (second) system. We now suppose at
each Lab that observers may make a measurement of either $\sigma_{z}$
(as performed by the Friends), or $S_{y}$ (as performed by the superobservers).
The Friends' measurement of $\sigma_{z}$ is modelled by a Hamiltonian
$H_{Am}$ and $H_{Bm}$ which leads to the coupling with the meter
modes. Hence, 
\begin{equation}
|\psi_{zz,m}\rangle=e^{-iH_{Am}t/\hbar}e^{-iH_{Bm}t/\hbar}|\psi_{zz}\rangle.\label{eq:ini-meter}
\end{equation}
We give an example of $H_{Am}$ in the Appendix. %
\begin{comment}
\begin{align*}
|+\rangle_{y,A} & =\frac{1}{\sqrt{2}}(|\alpha\rangle_{z}+i|-\alpha\rangle_{z})\\
|-\rangle_{y,A} & =\frac{1}{\sqrt{2}}(|-\alpha\rangle_{z}+i|\alpha\rangle_{z})\\
|\pm\alpha\rangle_{z,A} & =\frac{1}{\sqrt{2}}(|\pm\rangle_{y}-i|\mp\rangle_{y})\\
|\pm\beta\rangle_{z,B} & =\frac{1}{\sqrt{2}}(|\pm\rangle_{y}-i|\mp\rangle_{y})
\end{align*}
\end{comment}
\begin{comment}
\begin{align*}
U_{A}^{-1}|\alpha\rangle & =\frac{1}{\sqrt{2}}\left(|\alpha\rangle-i|-\alpha\rangle\right)\\
U_{A}^{-1}|-\alpha\rangle & =\frac{1}{\sqrt{2}}\left(|-\alpha\rangle-i|\alpha\rangle\right)
\end{align*}
\end{comment}
The state of the Labs after the Friends' measurements is given by
\begin{equation}
|\psi_{zz,mF}\rangle=e^{-iH_{AmF}t/\hbar}e^{-iH_{BmF}t/\hbar}|\psi_{zz,m}\rangle\label{eq:ini-f-meter}
\end{equation}
 which describes that the Friends have themselves become coupled
(entangled) with the meters. The coupling is given by interactions
$H_{AmF}$ and $H_{BmF}$.

We suppose the both Wigner superobservers measure $S_{y}$. The interactions
$H_{Am}$, $H_{Bm}$, $H_{AmF}$ and $H_{BmF}$ are reversed. The
superobservers perform the respective local unitary interactions $U_{A}^{-1}$
and $U_{B}^{-1}$ to change the measurement settings from $z$ to
$y$. After the unitary evolution corresponding to the measurement
interaction, the state of the system is
\begin{eqnarray}
|\psi_{yy}\rangle_{FR} & = & U_{A}^{-1}U_{B}^{-1}|\psi_{zz}\rangle_{FR}\nonumber \\
 & = & \frac{i}{2\sqrt{3}}\{-3i|\alpha\rangle_{y}|\beta\rangle_{y}-i|-\alpha\rangle_{y}|-\beta\rangle_{y}\nonumber \\
 &  & \ \ \ +|\alpha\rangle_{y}|-\beta\rangle_{y}-|-\alpha\rangle_{y}|\beta\rangle_{y}\}.\label{eq:cat-yy}
\end{eqnarray}
Each system is then coupled to meter $M$ (in the superobservers'
Labs), using interactions $H_{AM}$ and $H_{BM}$. A pointer measurement
is then made so that the superobservers may record the outcomes. We
see that there is a nonzero probability of $1/12$ that both observers
get outcomes $-1$ and $-1$ for measurements of $S_{y}$.

The alternative set-up is where the Friend of Lab $B$ measures $\sigma_{z}$
and superobserver $A$ measures $S_{y}$. After the reversals, the
unitary interaction $U_{y}$, the state is
\begin{eqnarray}
|\psi_{yz}\rangle_{FR} & = & U_{A}^{-1}|\psi_{zz}\rangle_{FR}\nonumber \\
 & = & \frac{e^{i\pi/4}}{\sqrt{6}}\{2|\alpha\rangle_{y}|-\beta\rangle_{z}+|-\alpha\rangle_{y}|\beta\rangle_{z}\nonumber \\
 &  & \ \ \ \ -i|\alpha\rangle_{y}|\beta\rangle_{z}\}.\label{eq:fryz}
\end{eqnarray}
The probability of getting $-1$ for lab $B$ and $-1$ for Lab $A$
on measurements of $\sigma_{z}$ and $S_{y}$ respectively is zero:
$P_{--|yz}=0$

The FR logic is as before: If from (\ref{eq:fryz}), Wigner superobserver
$A$ gives $-1$ for $S_{y}$, it is inferred from the state $|\psi_{yz}\rangle_{FR}$
that the Friend $B$ is $+1$ for $\sigma_{z}$. But if the Friend
$B$ gets the result $+1$ for $\sigma_{z}$, then one infers from
the original state $|\psi_{zz}\rangle_{FR}$ at time $t_{1}$ (Eq.
(\ref{eq:fr1})) that the Friend for system $A$ measured down for
$\sigma_{z}$.

Yet, considering state $|\psi_{zy}\rangle$ of (\ref{eq:fr1}) for
measurements of $\sigma_{z}$ at $A$ and $S_{y}$ at $B$, if $A$
was in the state $-1$ as measured by the Friend $A$, then there
is no possibility to get outcome $-1$ for $S_{y}$ at $B$. The probability
of obtaining $-1$ and $-1$ for measurements $\sigma_{z}$ and $S_{y}$
is zero. This is seen by considering the measurement of $\sigma_{z}$
at $A$ and $S_{y}$ at $B$: The final state after the reversals
and unitary rotation at $B$ is
\begin{eqnarray}
|\psi_{zy}\rangle_{FR} & = & U_{B}^{-1}|\psi_{zz}\rangle_{FR}\nonumber \\
 & = & \frac{e^{i\pi/4}}{\sqrt{3}}\{\frac{1}{\sqrt{2}}|\alpha\rangle_{z}(|-\beta\rangle_{y}-i|\beta\rangle_{y})\nonumber \\
 &  & \ \ \ +\sqrt{2}|-\alpha\rangle_{z}|\beta\rangle_{y}\}.\label{eq:catzy}
\end{eqnarray}
$P_{--|zy}=0$. This implies the impossibility to get both outcomes
of $S_{y}$ being $-1$ at $A$ and $B$, in contradiction to the
earlier quantum result: $P_{--|yy}=0$. This gives the paradox for
a situation where all the measurements, including those initially
made by the Friends, are distinguishing between two macroscopically
distinct states.\textcolor{blue}{}

\section{Analysis using macroscopic realism}

The macroscopic version of the paradox identifies two macroscopically
distinct states available to the systems at each time $t_{i}$. This
gives an inconsistency in the realistic perception of the same event
by the two observers, even where those events are based on measurements
of macroscopic qubits, which is puzzling. Here, we examine the definitions
of macroscopic realism carefully, showing that a \emph{deterministic
form of macroscopic realism} is indeed falsified.

\subsection{Deterministic macroscopic (local) realism falsified}

\textbf{\emph{Result 4.A: }}The violation of the Wigner-Bell inequality
for the cat-state version of the experiment implies failure of deterministic
macroscopic realism.

\emph{Proof:} The Wigner-Bell inequality (\ref{eq:observer-independence})\emph{
would} hold, if simultaneous predetermined values for both measurements
$\sigma_{y}$ and $\sigma_{z}$ are identifiable for both systems
$A$ and $B$, for the state $|\psi_{zz}\rangle_{WF}$ (Eq. (\ref{eq:state})
or (\ref{eq:state-1-2})). With this assumption, one would specify
hidden variables $\lambda_{z}^{A}$ and $\lambda_{y}^{A}$ that predetermine
the result of the measurement of $\sigma_{z}^{A}$ and $\sigma_{y}^{A}$
respectively, and hidden variables $\lambda_{z}^{B}$ and $\lambda_{y}^{B}$
that predetermine the result of the measurement of $\sigma_{z}^{B}$
and $\sigma_{y}^{B}$ respectively, should those measurements be performed
by the Friends (or superobservers). In the set-up, the possible results
for each $\lambda$ identify macroscopically distinct states for the
system, as evident by writing the state $|\psi\rangle_{WF}$ in the
respective bases, as in Eqs. (\ref{eq:state-1-2})-(\ref{eq:state-1-2-1-1-1}).

Following the definitions of macroscopic realism given in Refs. \citep{legggarg-1,manushan-bell-cat-lg,macro-bell-lg},
we then refer to the assumption of the simultaneous predetermined
variables as \emph{deterministic macroscopic realism} (dMR). The
assumption naturally includes that of Locality, but we may also specify
the assumption as \emph{deterministic macroscopic (local) realism}
to make this clear. Following the original proofs of the Bell inequalities
\citep{chsh,cshim-review-2}, the Bell-Wigner inequality follows based
on the assertion of the simultaneous variables $\lambda_{z}^{A}$,
$\lambda_{y}^{A}$, $\lambda_{z}^{B}$ and $\lambda_{y}^{B}$. The
premise dMR also specifies that the measurements $S_{z}$ and $S_{y}$
of the superobservers are determined by the hidden variables $\lambda_{z}$
and $\lambda_{y}$, at the respective site. The assumption of dMR
is therefore falsified by the violation of the Bell-Wigner inequality.

\subsection{Failure of deterministic macroscopic realism: the FR paradox}

\textbf{\emph{Result 4.B:}} The premise of dMR is also falsifiable
for the FR paradox, where the assertion applies to the state $|\psi_{zz}\rangle_{FR}$
(Eq. (\ref{eq:ini})).

\emph{Proof:} A Table of all possible values $\lambda_{x}$ and $\lambda_{y}$
for each system is constructed and the impossibility of the outcomes
predicted by quantum mechanics is evident. The Table is given in the
Appendix. In the Table, the dMR states giving a zero probability
for the result $-$, $-$ for the measurements $\sigma_{z}^{A}S_{y}^{B}$
($ZY$) and also a zero probability for getting $-$, $-$ for measurements
$S_{y}^{A}\sigma_{z}^{B}$ ($Y$$Z$) can be identified. There is
no possibility of an FR inconsistency, since those states for which
the result is also $-$, $-$ for measurements $S_{Y}^{A}S_{y}^{B}$
($Y$$Y$) is highlighted by the stars, but this gives a nonzero probability
of $+$, $+$ for $\sigma_{z}^{A}\sigma_{z}^{B}$ ($Z$, $Z$), which
is inconsistent with the initial state $|\psi_{FR}\rangle\equiv|\psi_{zz}\rangle$.
The initial state has a zero probability for $+,$$+$: Therefore
this prediction is not possible for the state (\ref{eq:fr1}). The
Table gives a falsification of dMR, if the predictions of quantum
mechanics are correct.

It is also possible to falsify dMR for the FR system, using the Bell-Wigner
inequality. The choice of measurement is of either $S_{y}$ or $\sigma_{z}$
at each site. Hence we write the Bell-Wigner inequality for the two
Labs, $L_{A}=A$ and $L_{B}=B$ as in (\ref{eq:observer-independence}),
in terms of the average correlations of the probability distributions,
where $A,B\in\{-1,+1\}$, \textcolor{black}{}
\begin{align}
S= & \left|\langle A_{z}B_{z}\rangle+\langle A_{y}B_{z}\rangle+\langle A_{z}B_{y}\rangle-\langle A_{y}B_{y}\rangle\right|\leq2.\label{eq:Bell-CHSH}
\end{align}
The assumption of dMR implies local hidden variables that predefine
the values for $A_{1},A_{2},B_{1}$ and $B_{2}$ to be $+1$ or $-1$,
implying the existence of the joint probability $p(A_{1},A_{2},B_{1},B_{2})$
whose marginals satisfy the Clauser--Horne--Shimony--Holt Bell
inequality (CHSH). For the FR paradox, there are $4$ possible preparation
states, $|\psi_{zz}\rangle_{FR}$, $|\psi_{yz}\rangle_{FR}$, $|\psi_{zy}\rangle_{FR}$
and $|\psi_{yy}\rangle_{FR}$. Identifying $A_{x}=\sigma_{z}^{A}$,
$A_{y}=S_{y}^{A}$, $B_{z}=\sigma_{z}^{B}$, $B_{y}=S_{y}^{B}$, we
evaluate from these states the following moments:
\begin{eqnarray}
\langle\sigma_{z}^{A}\sigma_{z}^{B}\rangle & = & -1/3\nonumber \\
\langle S_{y}^{A}\sigma_{z}^{B}\rangle & = & -2/3\nonumber \\
\langle\sigma_{z}^{A}S_{y}^{B}\rangle & = & -2/3\nonumber \\
\langle S_{y}^{A}S_{y}^{B}\rangle & = & 2/3\label{eq:FR-bW}
\end{eqnarray}
\textcolor{blue}{}which gives a value of $S=7/3$. The violation
of the inequality falsifies dMR. $\square$\textcolor{blue}{}

\section{weak macroscopic realism}

In this section, we show how consistency between macroscopic realism
and the quantum predictions of the Wigner friends gedanken experiment
can be obtained. Since deterministic macroscopic realism is falsified
by the experiment, it is necessary to define macroscopic realism in
a less strict sense, as given by a weaker (more minimal) assumption.
This motivates the premise of \emph{weak macroscopic realism} (wMR).
We explain how wMR can be consistent with the Wigner friend paradoxes.

\subsection{Definition of weak macroscopic realism}

\begin{figure}[t]
\begin{centering}
\par\end{centering}
\begin{centering}
\includegraphics[width=1\columnwidth]{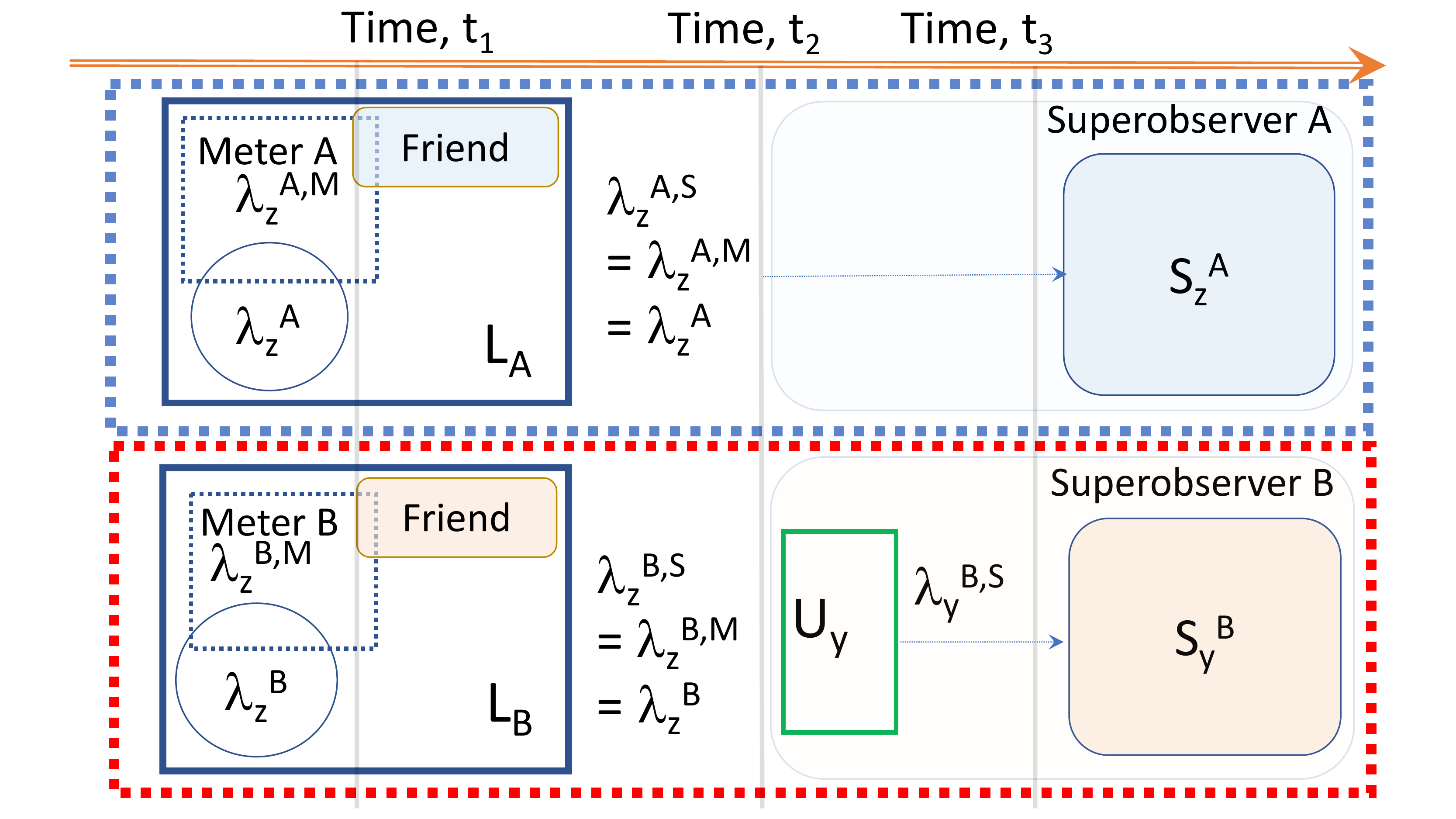}
\par\end{centering}
\caption{The weak macroscopic realism (wMR) model. The cat systems at time
$t_{1}$ are described by variables $\lambda_{z}^{A}$ and $\lambda_{z}^{B}$
that predetermine outcomes for measurements of $\sigma_{z}$. After
interaction with the cat systems, the meters are attributed values
$\lambda_{z}^{A,M}$ and $\lambda_{z}^{B,M}$ that predetermine the
outcome of the measurement on them by the Friends ($\lambda_{z}^{A,M}=\lambda_{z}^{A}$,
$\lambda_{z}^{B,M}=\lambda_{z}^{B}$). These variables predetermine
the outcomes ($\lambda_{z}^{A,S}$ and $\lambda_{z}^{B,S}$) \emph{if}
a measurement is made of $S_{z}$ by the superobservers $A$ or $B$,
at the time $t_{2}$. There is no change of measurement setting at
$A$. Hence $\lambda_{z}^{A,S}$ predetermines the outcome for $S_{z}^{A}$
by superobserver $A$ at time $t_{3}$, regardless of any unitary
$U_{y}$ at $B$. If superobserver $B$ performs a unitary interaction
$U_{y}$ to prepare the system for a (pointer) measurement $S_{y}$,
then the system at time $t_{3}$ is attributed a variable $\lambda_{y}^{B,S}$
that predetermines the outcome. It is now not necessarily true that
the value $\lambda_{z}^{B}$ predetermines the outcome of a future
measurement $S_{z}^{B}$ (this may depend on whether there is a further
unitary rotation at $A$). \label{fig:Exp-diagram-weakmr}}
\end{figure}

Weak macroscopic realism (wMR) is defined by the following two Assertions
(Figure \ref{fig:Exp-diagram-weakmr}) \citep{manushan-bell-cat-lg,ghz-cat}:

\textbf{\emph{Assertion wMR (1)}} \emph{There is realism for the system
prepared in the ``pointer superposition state'' i.e. for a system
prepared in the appropriate basis, for the pointer measurement}: The
system is regarded as having a definite predetermined value $\lambda_{\theta}$
(being $+1$ or $-1$) for the outcome $S_{\theta}$ \emph{after}
the unitary dynamics $U_{\theta}$ that occurs at a site, that determines
the choice of measurement setting $\theta$. This value $\lambda_{\theta}$
can be considered the value of the 'record' of the result of the measurement,
even if the final pointer stage of the measurement is not actually
carried out. We refer to $\lambda_{\theta}$ as the ``pointer value''.

\textbf{\emph{Assertion wMR (2)}} \emph{There is locality with respect
to this pointer value:} It is assumed that the value $\lambda_{\theta}$
predetermining the pointer measurement for $S_{\theta}$ is not changed
by spacelike separated events e.g. a unitary rotation $U_{\phi}$
at a separated Lab. $\square$

For measurements of the qubit value associated with the two coherent
states of system $A$, the final pointer stage of the measurement
corresponds to the determination of the sign of the quadrature phase
amplitude $X_{A}$. The premise wMR specifies a hidden variable
$\lambda_{\theta}^{A}$ for the outcome of this pointer measurement,
\emph{once} the dynamics $U_{\theta}^{A}$ associated with the choice
of measurement setting has occurred. This means that the predetermination
is with respect to one or other spin, $S_{z}$ or $S_{y}$, not (necessarily)
both simultaneously. For example, in Figure \ref{fig:Exp-diagram-weakmr},
the assertion applies to predetermine the outcome of $S_{z}$ at the
time $t_{2}$, and then to predetermine the outcome of $S_{y}$ at
the \emph{different} time $t_{3}$ (after the dynamics $U_{y}$).

As part of the definition of wMR, it is assumed that the value $\lambda_{\theta}^{A}$
predetermining the pointer measurement for $S_{\theta}^{A}$ at the
later time $t_{3}$ is not changed by the unitary rotation $U_{\phi}^{B}$
at $B$ (Figure \ref{fig:Exp-diagram-weakmr}). However, if there
is a further unitary rotation $U_{\theta'}^{A}$ at $A$, then a different
variable $\lambda_{\theta'}^{A}$ applies to the system at the later
time after the interaction $U_{\theta'}^{A}$, to predetermine the
result of the new pointer measurement $S_{\theta'}^{A}$. The wMR
postulate does not assert that this value is not affected by the unitary
rotation $U_{\phi}^{B}$, because here there are two unitary rotations
from the initial time of preparation ($t_{2}$). The postulate wMR
applies only to predetermine pointer measurements at the time after
the local unitary measurement setting operation, $U$.

\subsection{Achieving consistency: records and the breakdown of the Locality
assumption}

The assumptions of Brukner's Bell-Wigner inequality are: (1) Locality,
(2) Free choice, and (3) Observer-independent facts (a record from
a measurement should be a fact of the world that all observers can
agree on). The violation of the inequality implies at least one of
the assumptions breaks down.Here, we address which of these assumptions
can break down in a wMR-model and which records observers will agree
on.

\subsubsection{\emph{Records in a wMR model}}

We deduce which records the observers agree on in a wMR model from
the definition of wMR. We find:

\textbf{\emph{Result 5.B.1 (1):}} The Friends and superobservers agree
on the record for $\sigma_{z}$: From the definition of wMR, there
is predetermination of the value that would be the record of a measurement,
at a time $t$, \emph{after} the unitary interaction $U_{\theta}$
that determines the measurement setting $\theta$ for each Lab. Hence,
in the wMR model, a definite value $\lambda_{z}^{A}$ ($\lambda_{z}^{B}$)
predetermines the outcome of the Friend's spin measurement, $\sigma_{z}$,
at the Lab $A$ ($B$) (Figure \ref{fig:Exp-diagram-weakmr}). The
superobservers can make a corresponding measurement of $S_{z}$, through
various mechanisms which involve coupling to meters in the superobservers'
Labs, but do not involve a unitary interaction $U$ that gives a change
of a measurement basis. The pointer measurement made by the superobservers
can be regarded as a pointer measurement on the system of the Friends.
In the wMR model, there is a predetermined value $\lambda_{z}^{A,S}$
($\lambda_{z}^{B,S})$ for the outcome of the superobservers' measurement
$S_{z}^{A}$ ($S_{z}^{B}$), at the time $t_{4}$, and hence the wMR
model establishes that 
\[
\lambda_{z}^{A,S}=\lambda_{z}^{A},\ \lambda_{z}^{B,S}=\lambda_{z}^{B}.
\]
The value that gives the record of the Friends also gives the outcome
that would be obtained for the measurement made by the superobservers,
if they choose to measure the same spin, $S_{z}$, as the Friends
(Figure \ref{fig:Exp-diagram-weakmr}). There is agreement for these
records.

\textbf{\emph{Result 5.B.1 (2):}} There is consistency of records
between the Friends and superobservers, if only \emph{one} superobserver
measures a different spin component from that measured by the Friends.
In the wMR model, the inconsistency arises where the two superobservers
\emph{both} measure a different spin (e.g. $S_{y}$). This is due
to the unitary interactions $U$ that change the measurement settings.
We prove this in Section V.B.2 below.

\begin{figure}[t]
\begin{centering}
\par\end{centering}
\begin{centering}
\includegraphics[width=1\columnwidth]{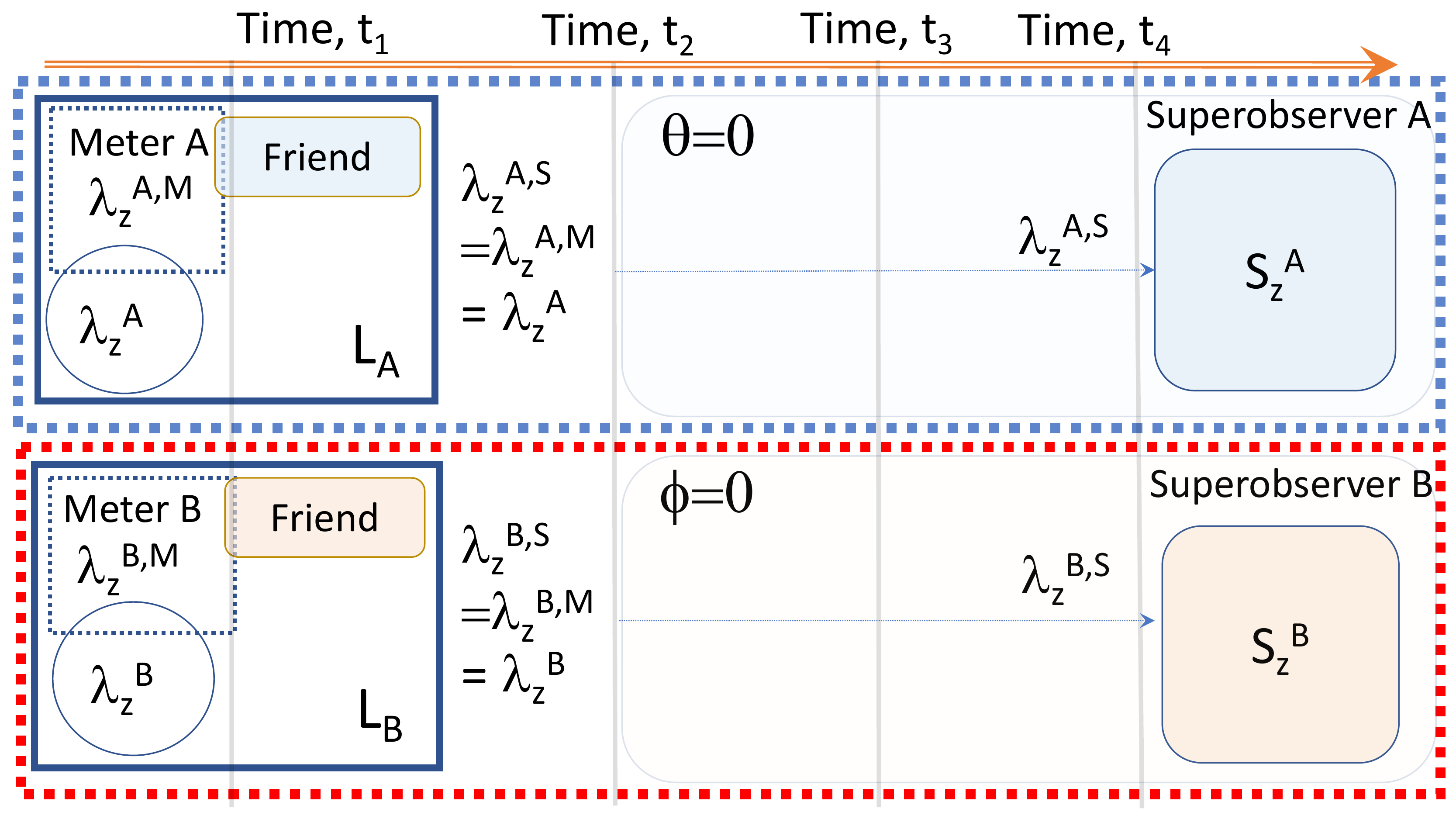}\vspace{0.05cm}
\par\end{centering}
\begin{centering}
\includegraphics[width=1\columnwidth]{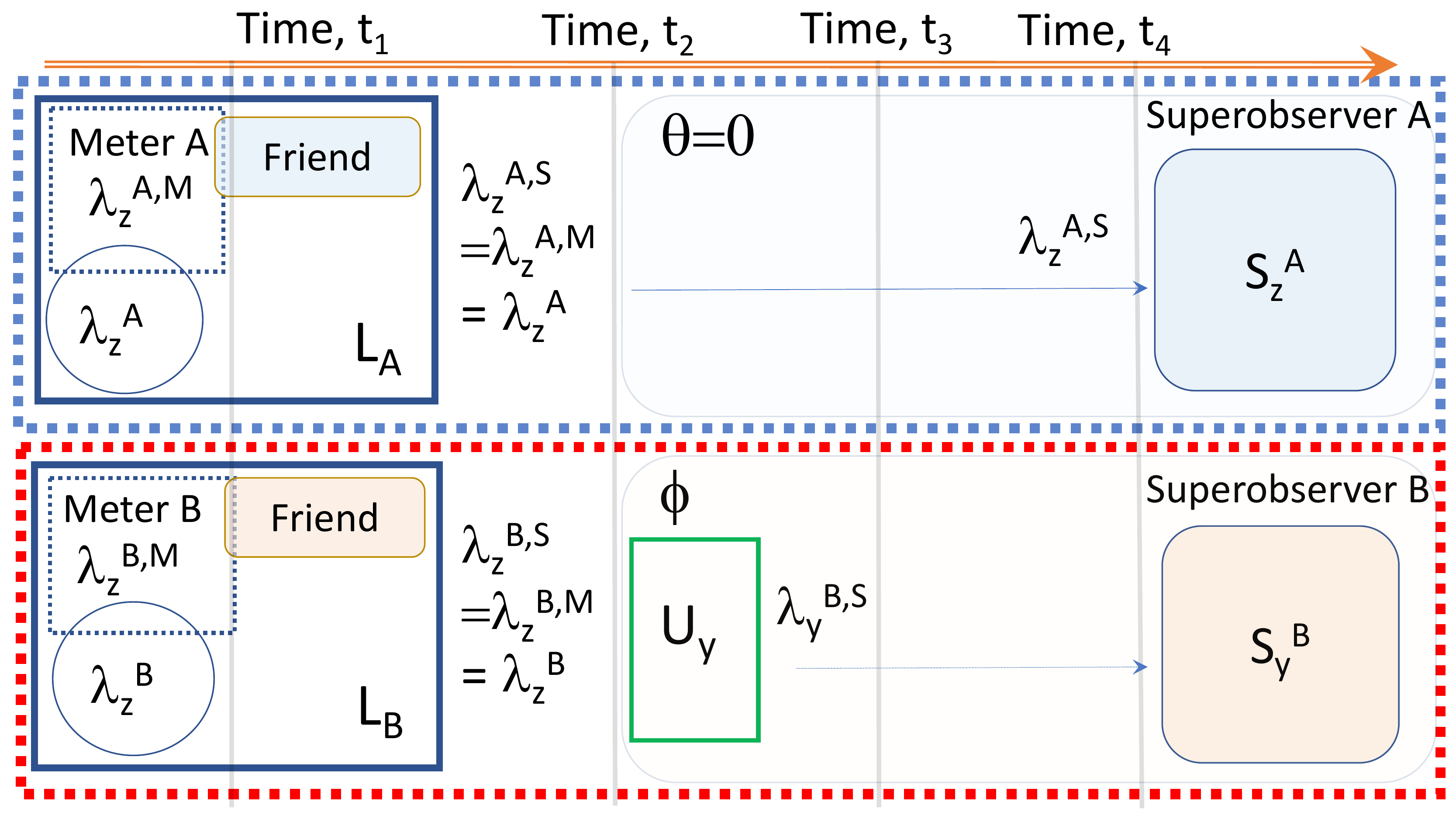}\vspace{0.05cm}
\par\end{centering}
\begin{centering}
\includegraphics[width=1\columnwidth]{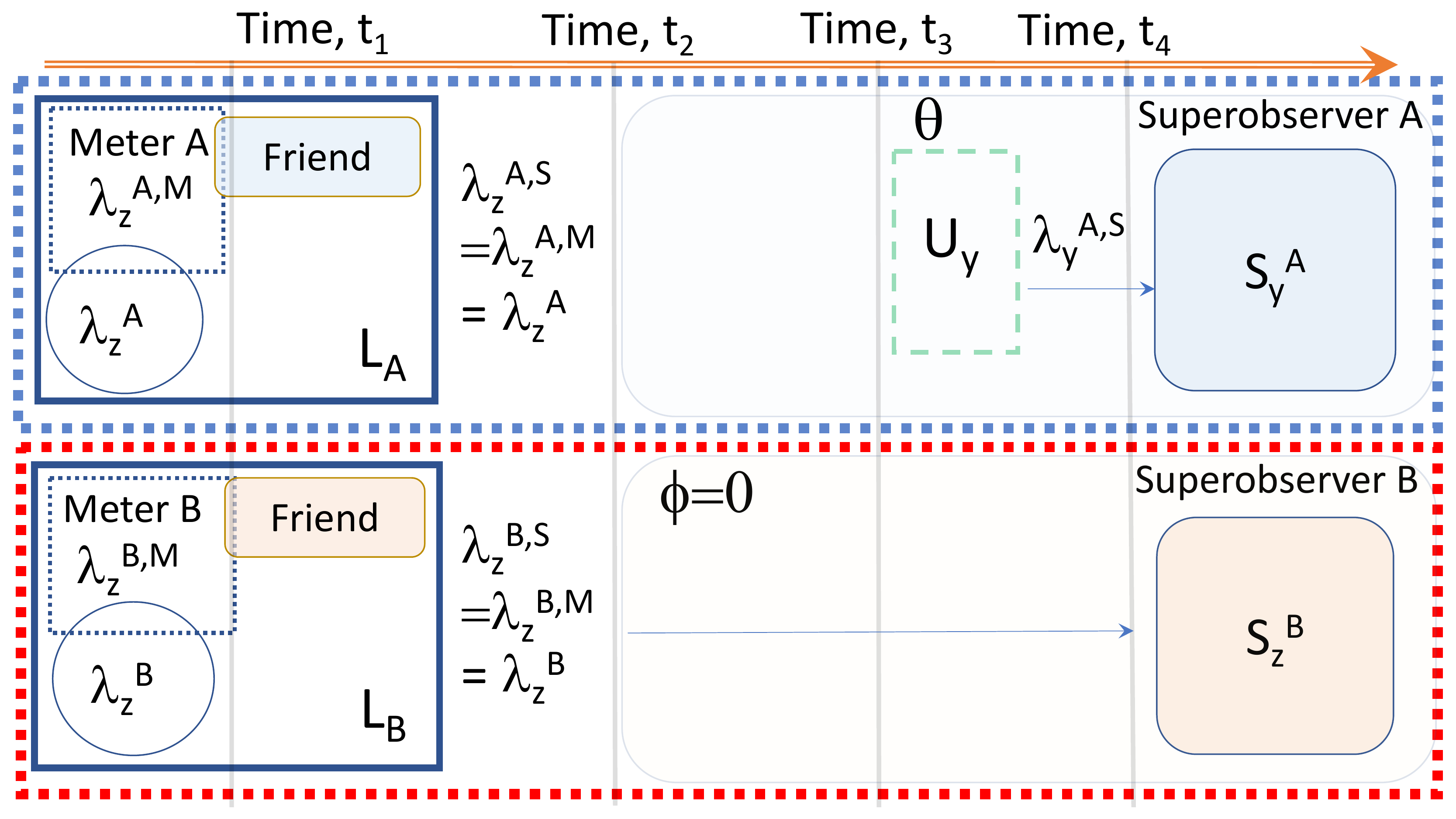}
\par\end{centering}
\caption{Resolving the paradox for consistency with weak macroscopic realism
(wMR): Partial Locality. According to wMR, measuring $\langle S_{z}^{A}S_{z}^{B}\rangle$,
$\langle S_{z}^{A}S_{y}^{B}\rangle$ and $\langle S_{y}^{A}S_{z}^{B}\rangle$,
which involve a unitary rotation at no more than one site, gives consistent
records between Friends and superobservers.  The systems at time
$t_{1}$ are prepared with respect to the basis for $\sigma_{z}$.
The premise wMR assigns variables $\lambda_{z}^{A}$, $\lambda_{z}^{B}$,
$\lambda_{z}^{A,M}$ and $\lambda_{z}^{B,M}$ to the Lab systems
so that $\langle S_{z}^{A}S_{z}^{B}\rangle=\langle\lambda_{z}^{A}\lambda_{z}^{B}\rangle$
(top). The superobserver $B$ may carry out a unitary interaction
$U_{y}$ to prepare the system with respect to the basis $Y$ (centre).
At time $t_{3}$, the system has a definite predetermined value for
$S_{y}^{B}$, given by $\lambda_{y}^{B,S}$, but the predetermination
of $S_{z}^{A}$ at site $A$ is unaffected. Hence $\langle S_{z}^{A}S_{y}^{B}\rangle=\langle\lambda_{z}^{A}\lambda_{y}^{B,S}\rangle$.
Similarly, $\langle S_{y}^{A}S_{z}^{B}\rangle=\langle\lambda_{y}^{A,S}\lambda_{z}^{B}\rangle$
(lower).\label{fig:Exp-diagram-track}}
\end{figure}

\begin{figure}[t]
\begin{centering}
\includegraphics[width=1\columnwidth]{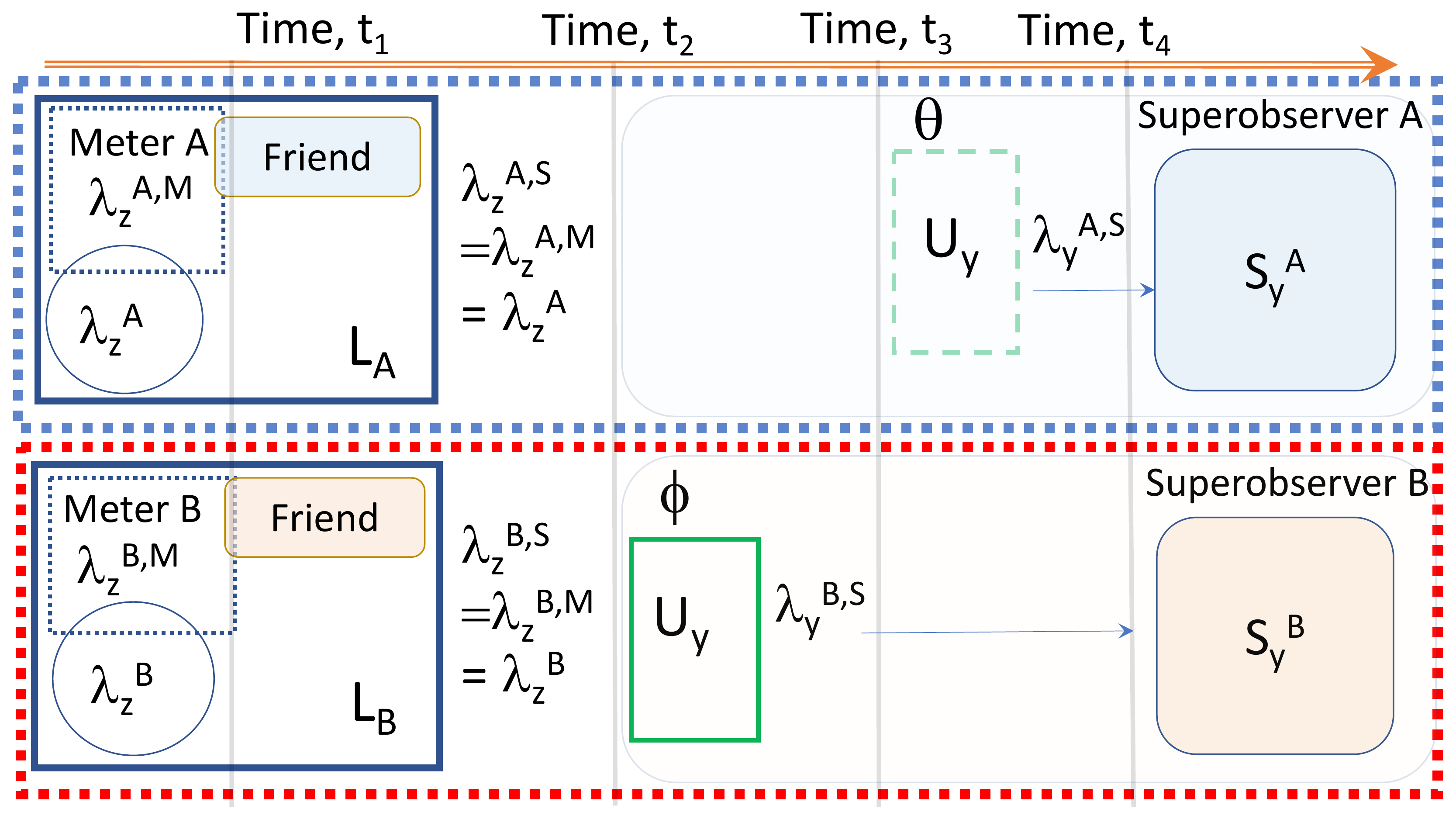}
\par\end{centering}
\caption{Resolving the paradox for consistency with weak macroscopic realism
(wMR): Locality need not hold and the paradox arises where there are
two rotations after the preparation, one at each site. The system
is prepared at time $t_{2}$ for pointer measurements of $S_{z}$
at each site. The premise of wMR assigns variables $\lambda$ to predetermine
the outcome after the unitary rotations that determine the measurement
settings, $\theta$ and $\phi$. This means that when there are two
unitary interactions $U_{y}$ as in the measurement of $\langle S_{y}^{A}S_{y}^{B}\rangle$
depicted in the diagram, the premise of wMR does not specify that
the value of $\lambda_{y}^{A,S}$ (defined after $\lambda_{y}^{B,S}$)
is independent of $\phi$, and the records are not necessarily consistent
according to the inequality (\ref{eq:observer-independence-2}).\label{fig:Exp-diagram-track-1}}
\end{figure}

\subsubsection{\emph{Results about Locality in a wMR model}}

The assumption of wMR implies only a partial locality. There is locality
with respect to the pointer values $\lambda_{z}$ defined after the
unitary interaction $U_{\theta}$. However, wMR does not imply Locality
in the full sense. It cannot be assumed that the future outcome of
$S_{y}$ at one Lab (as to be measured \emph{after} the unitary interaction
$U_{y}$ needed for the measurement setting) is independent of the
measurement choice $\phi$ occurring at the other Lab. Hence, the
observed violations of the Bell-Wigner inequality are not inconsistent
with wMR. We prove the following:

\textbf{\emph{Result 5.B.2: }}Weak macroscopic realism (wMR) does
not imply the Bell-CHSH inequality.

\emph{Proof: }This inequality is derived for the Wigner friend set-up
by noting the variables $\lambda_{z}^{A}$, $\lambda_{y}^{A}$, $\lambda_{z}^{B}$
and $\lambda_{y}^{B}$ have the values either $+1$ or $-1$, which
bounds the quantity 
\begin{equation}
S=\lambda_{z}^{A}\lambda_{z}^{B}+\lambda_{y}^{A}\lambda_{z}^{B}+\lambda_{z}^{A}\lambda_{y}^{B}-\lambda_{y}^{A}\lambda_{y}^{B}\label{eq:bell-3}
\end{equation}
so that $|S|\leq2$. The derivation is that these values exist in
any single run, so that the bound corresponds to that for the averages.
In the top diagram of Figure \ref{fig:Exp-diagram-track}, we see
we can assign the values $\lambda_{z}^{A}(t_{2})\equiv\lambda_{z}^{A}$
and $\lambda_{z}^{B}(t_{2})\equiv\lambda_{z}^{B}$ at time $t_{2}$,
so that we can put
\begin{equation}
\langle S_{z}^{A}S_{z}^{B}\rangle=\langle\lambda_{z}^{A}(t_{2})\lambda_{z}^{B}(t_{2})\rangle=\langle\lambda_{z}^{A}\lambda_{z}^{B}\rangle.\label{eq:bell-4}
\end{equation}
We then consider the centre diagram, and define $\lambda_{y}^{B}(t_{3})\equiv\lambda_{y}^{B}(t_{3}|\theta=0)$
as the value predetermining $S_{y}^{B}$, where there is no rotation
$U_{A}$ at $A$. According to wMR, the pointer measurement value
at $A$ is not affected by $U_{B}$, so that 
\begin{equation}
\langle S_{z}^{A}S_{y}^{B}\rangle=\langle\lambda_{z}^{A}\lambda_{y}^{B}(t_{3})\rangle=\langle\lambda_{z}^{A}\lambda_{y}^{B,S}\rangle.\label{eq:bell8}
\end{equation}
Similarly, we consider the lower diagram, and define $\lambda_{y}^{A}(t_{4})\equiv\lambda_{y}^{A}(t_{4}|\phi=0)$,
so that
\begin{equation}
\langle S_{y}^{A}S_{z}^{B}\rangle=\langle\lambda_{y}^{A}(t_{4})\lambda_{z}^{B}\rangle=\langle\lambda_{y}^{A,S}\lambda_{z}^{B}\rangle.\label{eq:bell9}
\end{equation}
The Figure \ref{fig:Exp-diagram-track-1} shows one way to measure
the moment $\langle S_{y}^{A}S_{y}^{B}\rangle$. The value of $\lambda_{y}^{B}(t_{3})$
determines $S_{y}^{B}$ independently of the future choice of $\theta$,
according to wMR, because the value of the pointer measurement is
specified at the time$t_{3}$ after the rotation $U_{y}$. We define
$\lambda_{y}^{A}(t_{4}|\phi\neq0)$ and $\lambda_{y}^{B}(t_{3})$,
and we can say 
\begin{equation}
\langle S_{y}^{A}S_{y}^{B}\rangle=\langle\lambda_{y}^{A}(t_{4}|\phi)\lambda_{y}^{B}(t_{3})\rangle=\langle\lambda_{y}^{A,S}\lambda_{y}^{B,S}\rangle.\label{eq:bell-10}
\end{equation}
However, from the postulate of wMR, we \emph{cannot} assume the value
$\lambda_{y}^{A}(t_{4})=\lambda_{y}^{A}(t_{4}|\phi)$ is independent
of $\phi$. This leads us to conclude consistency of values (records)
for the measurements carried out where there is no more than one unitary
rotation (as in Figure \ref{fig:Exp-diagram-track}), but not necessarily
where there are two changes of measurement setting, as in the measurement
of $\langle S_{y}^{A}S_{y}^{B}\rangle$ (Figure \ref{fig:Exp-diagram-track-1}).
$\square$

\section{Consistency of the quantum predictions with the two wMR assertions}

In this section, we \emph{explicitly} show that the quantum predictions
are consistent with the two assertions of the wMR premise, as stated
by the definition in Section V.B. The first assertion is that the
system prepared (after the unitary dynamics that determines the measurement
setting) for the pointer measurement has a predetermined outcome $\lambda$.
The second assertion is that this value is not altered by the dynamics
at a different site. We show the consistency by comparing the quantum
predictions with those of certain mixed states that give a particular
wMR model, thereby satisfying the wMR assertions.

We will illustrate with Figures for the FR paradox. Here, the system
is prepared at time $t_{1}$ in the state $|\psi_{zz}\rangle_{FR}$
given by (\ref{eq:ini}). It will be useful to depict the measurement
dynamics associated with the unitary rotations $U_{\theta}$ in terms
of the $Q$ function. The single-mode $Q$ function $Q(\alpha_{0})=\frac{1}{4\pi}|\langle\alpha_{0}|\psi\rangle|^{2}$
defines the quantum state $|\psi\rangle$ uniquely as a positive probability
distribution \citep{husimi}. The $Q$ function of the state $|\psi_{zz}\rangle_{FR}$
is $Q_{zz}(X_{A},P_{A},X_{B},P_{B})$, where
\begin{eqnarray}
{\normalcolor {\color{blue}{\normalcolor Q}_{{\normalcolor zz}}}} & {\color{blue}{\normalcolor =}} & {\normalcolor {\color{blue}{\normalcolor \frac{1}{\pi^{2}}|\langle\beta_{0},\alpha_{0}|\psi_{zz}\rangle|^{2}}}}\nonumber \\
 & {\color{blue}{\normalcolor =}} & \frac{e^{-|\alpha|^{2}-|\alpha_{0}|^{2}-|\beta|^{2}-|\beta_{0}|^{2}}}{3\pi^{2}}\Bigl\{{\normalcolor e^{-2\alpha X_{A}-2\beta X_{B}}}\nonumber \\
 & {\color{blue}} & +2\cosh(2\alpha X_{A}-2\beta X_{B})+2\cos(2\alpha P_{A}-2\beta P_{B})\nonumber \\
 &  & -2e^{-2\beta X_{B}}\sin(2\alpha P_{A})-2e^{-2\alpha X_{A}}\sin(2\beta P_{B})\Bigl\}.\nonumber \\
\label{eq:qzz}
\end{eqnarray}
Here, $\alpha_{0}=X_{A}+iP_{A}$, $\beta_{0}=X_{B}+iP_{B}$ and we
consider $\alpha$, $\beta$ to be real. The first two terms in brackets
give three distinct Gaussian peaks, corresponding to the three outcomes
for the joint spins, originating from $|\alpha\rangle|-\beta\rangle$,
$|-\alpha\rangle|\beta\rangle$ and $|-\alpha\rangle|-\beta\rangle$
in (\ref{eq:ini}). These terms constitute the $Q$ function of the
mixture $\rho_{zz}$ of the three states. The function $Q_{zz}$ has
three sinusoidal terms, which distinguish the superposition $|\psi_{zz}\rangle_{FR}$
from the mixture $\rho_{zz}$.

The $Q$ function corresponds to the anti-normally ordered operator
moments. We may compare with the probability distribution $P(\mathbf{X}_{A},\mathbf{X}_{B})$
for detecting outcomes $\mathbf{X}_{A}$ and $\mathbf{X}_{B}$ of
measurements of $\hat{X}_{A}$ and $\hat{X}_{B}$. While the peaks
of the marginal function $Q_{zz}(X_{A},X_{B})$ (defined by integrating
over $P_{A}$ and$P_{B}$) will show extra noise, this noise is at
the vacuum level. Here, we consider superpositions of macroscopically
distinct coherent states, as in $|\psi_{zz}\rangle_{FR}$ (Eq. (\ref{eq:ini})),
and the relevant spin measurements $\sigma_{\theta}$ or $S_{\theta}$
bin the amplitudes according to sign. To determine the spin outcomes,
it is only necessary to distinguish between the macroscopically distinct
components. Hence, the relative probabilities for the spin outcomes
are immediately evident from the solutions (and plots) of the marginal
$Q$ functions.

\subsection{Measuring $S_{z}^{A}$ and $S_{z}^{B}$: consistency with wMR}

We first show consistency of the predictions for $\langle\sigma_{z}^{A}\sigma_{z}^{B}\rangle$
and $\langle S_{z}^{A}S_{z}^{B}\rangle$ with a wMR model, thereby
verifying the first part of the definition of wMR. Consider the
system prepared at time $t_{1}$ in the state $|\psi_{zz}\rangle$.
The preparation is with respect to the $\sigma_{z}$ ($S_{z}$) ``pointer
basis'' at each Lab, so that a pointer measurement is all that is
needed to complete the measurement of $\sigma_{z}$ ($S_{z}$).

\textbf{\emph{Result}}\emph{ }\textbf{\emph{6.A:}} The system $|\psi_{zz}\rangle$
as prepared for the spin $z$ pointer measurements gives predictions
consistent with a wMR model.

\emph{Proof: }The essential feature of the proof is the comparison
between the predictions of the superposition as written in the pointer
basis with that of the corresponding \emph{mixed} state. With reference
to $|\psi_{zz}\rangle=c_{11}|\alpha\rangle|\beta\rangle+c_{12}|\alpha\rangle|-\beta\rangle+c_{21}|-\alpha\rangle|\beta\rangle+c_{22}|-\alpha\rangle|-\beta\rangle$,
the corresponding mixed state is
\begin{equation}
\rho_{zz}=\sum_{ij}|c_{ij}|^{2}\rho_{ij}\label{eq:mix-2}
\end{equation}
where $\rho_{11}=|\alpha\rangle|\beta\rangle\langle\alpha|\langle\beta|$,
$\rho_{12}=|\alpha\rangle|-\beta\rangle\langle\alpha|\langle-\beta|$,
$\rho_{21}=|-\alpha\rangle|\beta\rangle\langle-\alpha|\langle\beta|$
and $\rho_{22}=|-\alpha\rangle|-\beta\rangle\langle-\alpha|\langle-\beta|$.
The predictions of $|\psi_{zz}\rangle$ and $\rho_{zz}$ for the joint
probabilities of the pointer measurements $\sigma_{z}^{A}$ and $\sigma_{z}^{B}$
are identical. The premise of wMR asserts that hidden variables $\lambda_{z}^{A}$
and $\lambda_{z}^{B}$ are valid to predetermine the outcome of the
pointer measurements $\sigma_{z}^{A}$ and $\sigma_{z}^{B}$, respectively.
This interpretation holds for the mixed state $\rho_{zz}$, which
describes a system that is indeed \emph{in one or other} of the states
comprising the mixture, and hence describable by such variables $\lambda_{z}^{A}$
and $\lambda_{z}^{B}$ at the time $t_{1}$. Hence, since the predictions
for the \emph{pointer} measurement on $|\psi_{zz}\rangle_{FR}$ are
identical, a wMR model exists to describe the (pointer) predictions
for $|\psi_{zz}\rangle_{FR}$. $\square$

It is useful to visualize this result for the macroscopic system
by examining the $Q$ function, where one includes the meters. Consider
$|\psi_{zz}\rangle_{FR}$ given by (\ref{eq:ini}). The $Q$ function
for the state (\ref{eq:ini-meter}) where the meters are explicitly
included is similar to (\ref{eq:qzz}), but with four modes. The state
is expanded as
\begin{eqnarray}
|\psi_{zz,m}\rangle_{FR} & = & \frac{1}{\sqrt{3}}|\alpha\rangle_{z}|\gamma\rangle_{Am}|-\beta\rangle_{z}|-\gamma\rangle_{Bm}\nonumber \\
 &  & \ \ \ +\frac{1}{\sqrt{3}}|-\alpha\rangle_{z}|-\gamma\rangle_{Am}(|\beta\rangle_{z}|\gamma\rangle_{Bm}\nonumber \\
 &  & \ \ \ +i|-\beta\rangle_{z}|-\gamma\rangle_{Bm})\label{eq:ini-2-1}
\end{eqnarray}
where $|\gamma\rangle$, $|-\gamma\rangle$ are coherent states for
the meter of the Friend's systems. We take $\gamma$ as large and
real. The $Q$ function is $Q_{zz,m}=\frac{1}{\pi^{4}}|\langle\beta_{0},\alpha_{0},\gamma_{A},\gamma_{B}|\psi_{zz,m}\rangle_{FR}|^{2}$.
Defining the complex variables $\gamma_{A}=X_{\gamma A}+iP_{\gamma A}$
for meter mode $Am$ of system $A$, and $\gamma_{B}=X_{\gamma B}+iP_{\gamma_{B}}$
for meter mode $Bm$ for system $B$, we find ($\alpha_{0}=X_{A}+iP_{A}$,
$\beta_{0}=X_{B}+iP_{B}$)
\begin{align}
Q_{zz,m} & =\frac{e^{-|\alpha|^{2}-|\beta|^{2}-2|\gamma|^{2}}}{3\pi^{4}}e^{-|\alpha_{0}|^{2}-|\beta_{0}|^{2}-|\gamma_{A}|^{2}-|\gamma_{B}|^{2}}\nonumber \\
 & \,\,\,\,\Bigl\{ e^{-2\alpha X_{A}-2\gamma X_{\gamma A}}e^{-2\beta X_{B}-2\gamma X_{\gamma B}}\nonumber \\
 & \,+2\cosh(2\alpha X_{A}+2\gamma X_{\gamma A}-2\beta X_{B}-2\gamma X_{\gamma B})\nonumber \\
 & \,+2\cos(2\alpha P_{A}+2\gamma P_{\gamma A}-2\beta P_{B}-2\gamma P_{\gamma B})\nonumber \\
 & \,-2e^{-2\beta X_{B}-2\gamma X_{\gamma B}}\sin(2\alpha P_{A}+2\gamma P_{\gamma A})\nonumber \\
 & -2e^{-2\alpha X_{A}-2\gamma X_{\gamma A}}\sin(2\beta P_{B}+2\gamma P_{\gamma B})\Bigl\}.\label{eq:qzz_4Mode-1}
\end{align}
The last three terms decay as $e^{-\gamma^{2}}$, and so for large
$\gamma$, the solution is 
\begin{align}
Q_{zz,m} & =\frac{e^{-P_{A}^{2}-P_{B}^{2}-P_{\gamma A}^{2}-P_{\gamma B}^{2}}}{3\pi^{4}}\nonumber \\
 & \,\,\,\,\Bigl\{ e^{-(X_{A}+\alpha)^{2}}e^{-(X_{B}+\beta)^{2}}e^{-(X_{\gamma A}+\gamma)^{2}}e^{-(X_{\gamma B}+\gamma)^{2}}\nonumber \\
 & +e^{-(X_{A}-\alpha)^{2}}e^{-(X_{B}+\beta)^{2}}e^{-(X_{\gamma A}-\gamma)^{2}}e^{-(X_{\gamma B}+\gamma)^{2}}\nonumber \\
 & +e^{-(X_{A}+\alpha)^{2}}e^{-(X_{B}-\beta)^{2}}e^{-(X_{\gamma A}+\gamma)^{2}}e^{-(X_{\gamma B}-\gamma)^{2}}\Bigl\}.\nonumber \\
\label{eq:qzz_4Mode-1-1-2}
\end{align}
The final meter (pointer) measurement corresponds to the measurement
of the meter quadrature amplitudes $X_{\gamma A}$ and $X_{\gamma B}$.
The marginal $Q(X_{\gamma A},X_{\gamma B})$ of that describes the
distribution for the measured meter outputs (as measured by the Friends)
is found by integrating over all system variables as well as $P_{\gamma A}$
and $P_{\gamma B}$: We find
\begin{align}
Q_{zz,m}(X_{\gamma A},X_{\gamma B}) & =\frac{1}{3\pi}\Bigl\{ e^{-(X_{\gamma A}+\gamma)^{2}-(X_{\gamma B}+\gamma)^{2}}\nonumber \\
 & +e^{-(X_{\gamma A}-\gamma)^{2}}e^{-(X_{\gamma B}+\gamma)^{2}}\nonumber \\
 & +e^{-(X_{\gamma A}+\gamma)^{2}}e^{-(X_{\gamma B}-\gamma)^{2}}\Bigl\}.\label{eq:Qmarg_4mode-1-1-1-1}
\end{align}
The three Gaussians are well-separated peaks, which represent the
three distinct sets of outcomes, as expected from the components of
$|\psi_{zz}\rangle_{FR}$ (Eq. (\ref{eq:ini})).

The function $Q_{zz,m}(X_{\gamma A},X_{\gamma B})$ gives the probabilities
for detection of each component, for the measurement on the meter
made by the Friends. We see this  corresponds to $Q_{zz}(X_{A},X_{B})$
of the\emph{ }mixed state $\rho_{zz}$ (Eq. (\ref{eq:mix-2}))
\begin{equation}
\rho_{zz}=\frac{1}{3}(\rho_{1}+\rho_{2}+\rho_{3})\label{eq:mix}
\end{equation}
where $\rho_{1}=|\alpha\rangle|-\beta\rangle\langle\alpha|\langle-\beta|$,
$\rho_{2}=|-\alpha\rangle|\beta\rangle\langle-\alpha|\langle\beta|$
and $\rho_{3}=|-\alpha\rangle|-\beta\rangle\langle-\alpha|\langle-\beta|$,
once we put $X_{\gamma A}=X_{A}$ and $X_{\gamma B}=X_{B}$, in (\ref{eq:Qmarg_4mode-1-1-1-1}).
This is expected, since the meter outcomes are a measurement of the
system amplitudes, $\hat{X}_{A}$ and $\hat{X}_{B}$. 

\begin{figure}[t]
\begin{centering}
\par\end{centering}
\begin{centering}
\includegraphics[width=1\columnwidth]{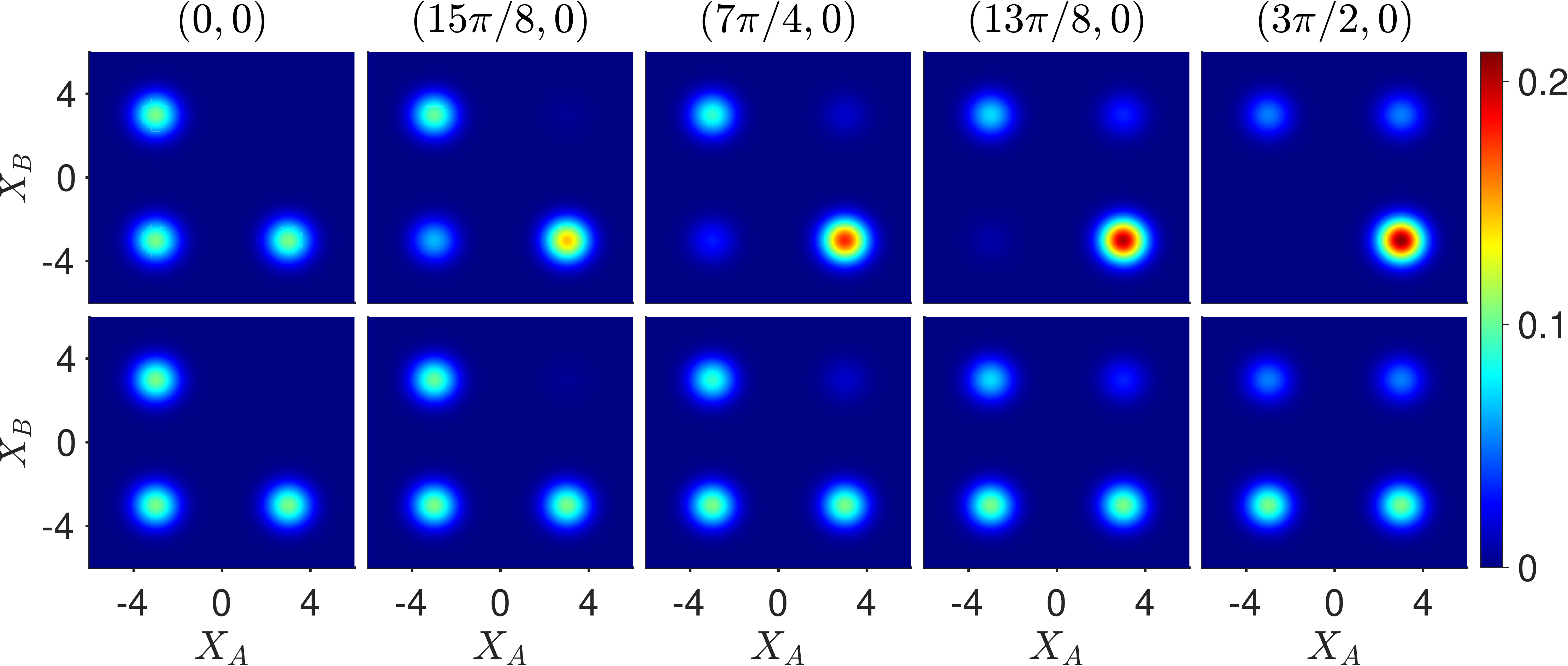}
\par\end{centering}
\caption{\textcolor{brown}{}Contrasting the dynamics for the superposition
state $|\psi_{zz}\rangle_{FR}$ and the mixed state $\rho_{zz}$.
The states are initially indistinguishable but become macroscopically
distinguishable under evolution. (top sequence, from left to right)
 We show contour plots of the marginal function $Q(X_{A},X_{B})$
at the times $(t_{a},t_{b})$, after the system has undergone a local
evolution under $H_{NL}$ (Eq. (\ref{eq:mint})) for a time $t_{a}$
and $t_{b}$ in each Lab. The system begins in $|\psi_{zz}\rangle_{FR}$
(Eq. (\ref{eq:ini})) (top, far left) and evolves to $|\psi_{yz}\rangle_{FR}$
(Eq. (\ref{eq:fryz}) (top, far right).  (lower sequence, from
left to right) The system begins in $\rho_{zz}$ (Eq. (\ref{eq:mix-2}))
and is evolved as for the top plots. Here, $\alpha=3$, $\Omega=1$.
\textcolor{blue}{\label{fig:q-function-catstatezz-yz-1}}\textcolor{red}{}\textcolor{blue}{}}
\end{figure}

We may further compare the distribution (\ref{eq:Qmarg_4mode-1-1-1-1}),
which describes the final outputs of the meter-measurements made by
the Friends, with that of the marginal $Q$ function
\begin{align}
Q_{zz}(X_{A},X_{B}) & =\frac{e^{-|\alpha|^{2}-X_{A}^{2}-|\beta|^{2}-X_{B}^{2}}}{3\pi}\Bigl\{{\normalcolor e^{-2\alpha X_{A}-2\beta X_{B}}}\nonumber \\
 & \left.+2\cosh(2\alpha X_{A}-2\beta X_{B}){\color{black}+2e^{-\alpha^{2}-\beta^{2}}}\right\} \label{eq:Qmarg-sup-2}
\end{align}
obtained directly from the superposition $|\psi_{zz}\rangle_{FR}$
(Eq. (\ref{eq:ini})). This is derived from $Q_{zz}$ (Eq. (\ref{eq:qzz})),
by integrating over $P_{A}$ and $P_{B}$. This function corresponds
to that of the systems $A$ and $B$ \emph{prior} to coupling to the
meter, and gives an alternative way to model the measurement by the
Friends. We see that for large $\alpha=\beta$, the last term vanishes,
which gives the result for $Q_{zz}(X_{A},X_{B})$ identical to (\ref{eq:Qmarg_4mode-1-1-1-1}),
on replacing $X_{\gamma A}$ and $X_{\gamma B}$ with $X_{A}$ and
$X_{B}$. This implies that in fact for the cat state where $\alpha$
and $\beta$ are large, the distribution for the outcomes of the \emph{pointer
measurement} on the superposition is indistinguishable from that for
the outcomes of the pointer measurement made on the mixed state.
This is evident from Figure \ref{fig:q-function-catstatezz-yz-1}.
The function $Q_{zz}(X_{A},X_{B})$ of (\ref{eq:Qmarg-sup-2}) is
plotted in Figure \ref{fig:q-function-catstatezz-yz-1} (far left
top snapshot) and is indistinguishable from (\ref{eq:Qmarg_4mode-1-1-1-1})
of the mixture $\rho_{zz}$ (far left lower snapshot, where $X_{\gamma A}$
and $X_{\gamma B}$ are labelled $X_{A}$ and $X_{B}$).

In summary, the $Q$ function solutions for the cat and meter states
give a convincing illustration of Result 6.A, that there is consistency
with wMR for $\langle\sigma_{z}^{A}\sigma_{z}^{B}\rangle$, the measurements
made by the Friends. The marginal $Q$ functions $Q_{zz}(X_{\gamma A},X_{\gamma B})$
and $Q_{zz}(X_{A},X_{B})$ for $|\psi_{FR}\rangle=|\psi\rangle_{zz}$
at time $t_{1}$ are indistinguishable from those of $\rho_{zz}$.
Hence, the predictions for the \emph{pointer} measurements of $\sigma_{z}^{A}$
and $\sigma_{z}^{B}$ on the system prepared in $|\psi_{FR}\rangle=|\psi\rangle_{zz}$
at time $t_{1}$ are consistent with a wMR model. We see however from
Figure \ref{fig:q-function-catstatezz-yz-1} that with the appropriate
evolution, despite that the distinction between the $Q$ function
for $|\psi_{FR}\rangle=|\psi\rangle_{zz}$ and $\rho_{zz}$ decays
with $e^{-\alpha^{2}}$, the evolved states show a consistent macroscopic
difference, even as $\alpha\rightarrow\infty$.

So far, the analysis concerns the measurements $\langle\sigma_{z}^{A}\sigma_{z}^{B}\rangle$
made by the Friends. We now consider $\langle S_{z}^{A}S_{z}^{B}\rangle$
as measured by the superobservers. If the system-meter state given
by $|\psi_{zz,m}\rangle$ (Eq. (\ref{eq:ini-f-meter})) is coupled
to the Friends, then the final state being written as
\begin{eqnarray}
|\psi_{zz,mF}\rangle & = & \frac{1}{\sqrt{3}}|\alpha\rangle_{z}|\gamma\rangle_{AmF}|-\beta\rangle_{z}|-\gamma\rangle_{BmF}\nonumber \\
 &  & \ \ \ +\frac{1}{\sqrt{3}}|-\alpha\rangle_{z}|-\gamma\rangle_{AmF}(|\beta\rangle_{z}|\gamma\rangle_{BmF}\nonumber \\
 &  & \ \ \ +i|-\beta\rangle_{z}|-\gamma\rangle_{BmF})\label{eq:ini-2-1-3}
\end{eqnarray}
where $|\pm\gamma\rangle_{AmF}=|\gamma\rangle_{Am}|F\rangle_{A}$
and $|\pm\gamma\rangle_{BmF}=|\gamma\rangle_{Bm}|F\rangle_{B}$ represent
the combined states of the meter and Friend in each Lab. The measurements
$S_{z}$ made by the superobservers correspond to measurements on
the system in a state of type (\ref{eq:ini-2-1}), except that the
meter systems are further coupled to second larger meters (e.g. the
Friends). Since \emph{only a pointer measurement} is necessary to
complete the measurement of $S_{z}$, the final distribution $Q_{zz}(X_{\gamma FA},X_{\gamma FB})$
for the outcomes of the superobservers, found after integration over
the unmeasured variables, is identical to (\ref{eq:Qmarg_4mode-1-1-1-1}),
the distribution for the mixed state $\rho_{zz}$ (once we put $X_{\gamma FA}=X_{A}$
and $X_{\gamma FB}=X_{B}$). Hence by the same argument as above,
the distribution and predictions for the final outcomes $S_{z}^{A}$
and $S_{z}^{B}$ measured by the superobservers are consistent with
variables $\lambda_{z}^{A,S}$ and $\lambda_{z}^{B,S}$, giving consistency
with wMR.

Moreover, we prove consistency with Result 5.B.1 (1) of wMR, that
the variables $\lambda_{z}^{A}$ and $\lambda_{z}^{B}$ predetermining
the outcomes of $\sigma_{z}^{A}$ and $\sigma_{z}^{B}$ for the Friends
\emph{are equal} to those ($\lambda_{z}^{A,S}$ and $\lambda_{z}^{B,S}$)
predetermining the outcomes of $S_{z}^{A}$ and $S_{z}^{B}$ for the
superobservers (Figure \ref{fig:Exp-diagram-weakmr}):
\begin{align}
\lambda_{z}^{A,S} & =\lambda_{z}^{A,m}=\lambda_{z}^{A}\nonumber \\
\lambda_{z}^{B,S} & =\lambda_{z}^{B,m}=\lambda_{z}^{B}.\label{eq:hv}
\end{align}
We follow the arguments above to note that the result $Q_{zz}(X_{\gamma FA},X_{\gamma FB})$
will be indistinguishable from that obtained if  the system had been
prepared in the mixture $\rho_{zz}$ at time $t_{1}$, and then measured
by the Friends and/ or superobservers. Such measurements on $\rho_{zz}$
will satisfy (\ref{eq:hv}). To prove this, consider the system prepared
in $\rho_{zz}$. Here, one can assign variables $\lambda_{z}^{A}$
and $\lambda_{z}^{B}$ which indicate the system is \emph{in} one
of the three states $\rho_{i}$ (of Eq. \ref{eq:mix-2}). This implies
predetermined outcomes for the spins $\sigma_{z}^{A}$ and $\sigma_{z}^{B}$,
consistent with wMR. If the Friends make a measurement on this system,
the solution is given precisely by (\ref{eq:qzz_4Mode-1-1-2}). The
combined system after the coupling to the meters is the correlated
mixed state
\begin{equation}
\rho_{mix,m}=\frac{1}{3}(\rho_{1}\rho_{\gamma1}+\rho_{2}\rho_{\gamma2}+\rho_{3}\rho_{\gamma3})\label{eq:mix6-1}
\end{equation}
where $\rho_{\gamma i}$ is the state of the meters. For the mixture,
it is valid to say that if the system were in the state $\rho_{i}$
at time $t_{1}$, then the meter after the coupling is in state $\rho_{\gamma i}$.
Here, because the combined system is a mixed state, one can assign
variables $\lambda_{z}^{A,m}$ and $\lambda_{z}^{B,m}$ to the meters,
these variables predetermining the outcomes of the measurements on
the meter. For the mixed state, the meter variables are correlated
with $\lambda_{z}^{A}$ and $\lambda_{z}^{B}$, those of the systems.
The outcomes of the Friend's measurements on the meters indicates
the values of the $\lambda_{z}^{A}$ and $\lambda_{z}^{B}$. Hence,
we put $\lambda_{z}^{A,m}=\lambda_{z}^{A}$ and $\lambda_{z}^{B,m}=\lambda_{z}^{B}$.

We then consider the system-meter-Friend state $|\psi_{zz,mF}\rangle$
given by (\ref{eq:ini-2-1-3}). On the other hand, if the system in
the mixed state $\rho_{mix,m}$ is coupled to another set of meters
(the Friends), then the system is described by
\begin{equation}
\rho_{mix,mF}=\frac{1}{3}(\rho_{1}\rho_{\gamma F1}+\rho_{2}\rho_{\gamma F2}+\rho_{3}\rho_{\gamma F3})\label{eq:mix7}
\end{equation}
which is a mixture of the three components in (\ref{eq:ini-2-1-3}),
$\rho_{\gamma Fi}$ being a state of the meters and Friends. As above,
the relevant marginal $Q$ distributions for $|\psi_{zz,mF}\rangle$
and $\rho_{mix,mF}$ that give the predictions for the pointer measurements
(the outcome for spin $Z$) are indistinguishable. One may assign
variables to the system (\ref{eq:mix7}), these variables predetermining
the outcome of the superobservers' measurements, so that (\ref{eq:hv})
holds. Hence, for the system originally prepared in $\rho_{zz}$,
the distributions and predictions for the measurements $\sigma_{z}$
made by the Friends and $S_{z}$ made by the superobservers are consistent
with (\ref{eq:hv}): Since these distributions and predictions are
indistinguishable from those for the system prepared in $|\psi_{zz}\rangle_{FR}$,
we conclude there is consistency of the predictions of the moment
$\langle S_{z}^{A}S_{z}^{B}\rangle$ with (\ref{eq:hv}). This implies
measurement by the superobservers if they measure $S_{z}$ will be
consistent with the records $\lambda_{z}^{A}$ and $\lambda_{z}^{B}$
obtained by the Friends.

\subsection{Measuring $S_{y}$: consistency of the unitary dynamics with wMR}

We next examine the measurements needed for the moments $\langle S_{y}^{A}S_{z}^{B}\rangle$,
$\langle S_{z}^{A}S_{y}^{B}\rangle$ and $\langle S_{y}^{A}S_{y}^{B}\rangle$.
For $\langle S_{y}^{A}S_{z}^{B}\rangle$ and $\langle S_{z}^{A}S_{y}^{B}\rangle$,
the system is prepared for the pointer measurement $S_{z}$ at the
time $t_{1}$ in one of the Labs, but a unitary rotation $U_{y}$
needs to be applied in the other Lab (Figure \ref{fig:Exp-diagram-track}).
The premise of wMR asserts a value $\lambda_{z}$ which predetermines
the outcome for $S_{z}$. This value applies to the system from the
time $t_{1}$, and is unaffected by the dynamics $U_{y}$ at the other
Lab. In this section, we show consistency with this assertion.

\textbf{\emph{Result 6.B (1):}} Consider a system prepared in the
pointer basis, which we choose to be spin $z$. The predictions where
there is a single further rotation $U$ in one of the Labs will be
consistent with wMR.

\emph{Proof:} We are considering a state of the type 
\begin{equation}
|\psi_{zz}\rangle=a_{+}|\beta\rangle|\psi\rangle_{A+}+a_{-}|-\beta\rangle|\psi\rangle_{A-}\label{eq:state-zz}
\end{equation}
where $|\psi\rangle_{A+}=c_{+}|\alpha\rangle+c_{-}|-\alpha\rangle$
and $|\psi\rangle_{A-}=d_{+}|\alpha\rangle+d_{-}|-\alpha\rangle$,
for probability amplitudes $a_{\pm}$, $c_{\pm}$ and $d_{\pm}$.
After a unitary rotation $U_{A}$ at $A$,
\begin{equation}
|\psi_{zz}(t)\rangle=a_{+}|\beta\rangle U_{A}(t)|\psi\rangle_{A+}+a_{-}|-\beta\rangle U_{A}(t)|\psi\rangle_{A-.}\label{eq:state-uni}
\end{equation}
The rotations give solutions of the form
\begin{eqnarray}
|\psi_{zz}(t)\rangle & = & a_{+}|\beta\rangle(c_{1}(t)|\alpha\rangle+c_{2}(t)|-\alpha\rangle)\nonumber \\
 &  & +a_{-}|-\beta\rangle(d_{1}(t)|\alpha\rangle+d_{2}(t)|-\alpha\rangle).\label{eq:state-dyn}
\end{eqnarray}
We first show that the predictions are indistinguishable from those
of the mixture 
\begin{equation}
\rho_{mix,B}=p_{+}|\beta\rangle\langle\beta|\rho_{A+}+p_{-}|-\beta\rangle\langle-\beta|\rho_{A-}\label{eq:rho-mix-b}
\end{equation}
where $p_{+}=|a_{+}|^{2}$ and $p_{-}=|a_{-}|^{2}$, which becomes
\begin{eqnarray}
\rho_{mix,B} & = & p_{+}|\beta\rangle\langle\beta|U_{A}\rho_{A+}U_{A}^{\dagger}\nonumber \\
 &  & +p_{-}|-\beta\rangle\langle-\beta|U_{A}\rho_{A-}U_{A}^{\dagger}.\label{eq:rho-mixbua}
\end{eqnarray}
On expansion, it is straightforward to show that the measurable probabilities
for the final mixed state are $|c_{1}(t)|^{2}$, $|c_{2}(t)|^{2}$,
$|d_{1}(t)|^{2}$ and $d_{2}(t)|^{2}$, identical to those of the
evolved state $|\psi_{zz}(t)\rangle$.

The second part of the proof is to show equivalence to wMR. Here,
there is preparation for the pointer measurement $S_{z}^{B}$ and
no further unitary dynamics occurs at Lab $B$. Weak macroscopic realism
implies a predetermined value $\lambda_{z}^{B}$ for the result of
$S_{z}^{B}$, and that this value is not affected by the unitary dynamics
$U_{y}^{A}$ that occurs in Lab $A$. For the mixture $\rho_{mix,B}$,
the system $B$ is \emph{in} one or other of the states, $\rho_{1}=|\beta\rangle\langle\beta|$
or $\rho_{2}=|-\beta\rangle\langle-\beta|$. As $\beta\rightarrow\infty$,
each of these states gives a definite outcome, $+1$ or $-1$ respectively,
for $\sigma_{z}^{B}$. This implies that the system $B$ at the initial
time is in a state with a predetermined value $\lambda_{z}^{B}$ for
the outcome of $\sigma_{z}^{B}$ and $S_{z}^{B}$. Any operations
by the superobserver in Lab $A$ are local. The system prepared in
$\rho_{mix,B}$ remains in a state with the definite value $\lambda_{z}$
for $\sigma_{z}^{B}$, throughout the dynamics. The dynamics for
the system prepared in $|\psi_{zz}\rangle$ under the evolution $U_{y}^{A}$
is \emph{indistinguishable} from that for $\rho_{mix-B}$. That dynamics
is therefore consistent with wMR. $\square$

\textbf{\emph{Result 6.B (2):}} The dynamics for the superposition
$|\psi_{zz}\rangle$ and the mixed state $\rho_{mix,B}$ can diverge,
if there are rotations $U$ and $U$ at both sites.

\emph{Proof:} This is easy to show, on expansion. We will also prove
this by example.

\subsubsection{Dynamics of the change of measurement setting}

We illustrate Result 6.B by examining the dynamics of the measurements.
To measure $S_{y}$, the superobserver must first reverse the coupling
of the system to the Friend and meter. The superobserver then performs
a local unitary rotation $U_{y}$, to change the measurement setting
from $x$ to $y$. This occurs over the timescale associated with
$U_{y}$. Following that, a pointer measurement occurs, by coupling
to a second meter in the superobserver's Lab, thereby completing the
measurement of $S_{y}$. We focus on the unitary dynamics $U_{y}$,
and assume the decoupling from the Friend-meter has been performed
by the time $t_{2}$ (Figures \ref{fig:Exp-diagram-track} and \ref{fig:Exp-diagram-track-1}).
As outlined in Section III, we assume that the reversal takes place
at both Labs, even where one superobserver may opt to measure $S_{z}$.

\begin{figure}[t]
\begin{centering}
\par\end{centering}
\begin{centering}
\includegraphics[width=1\columnwidth]{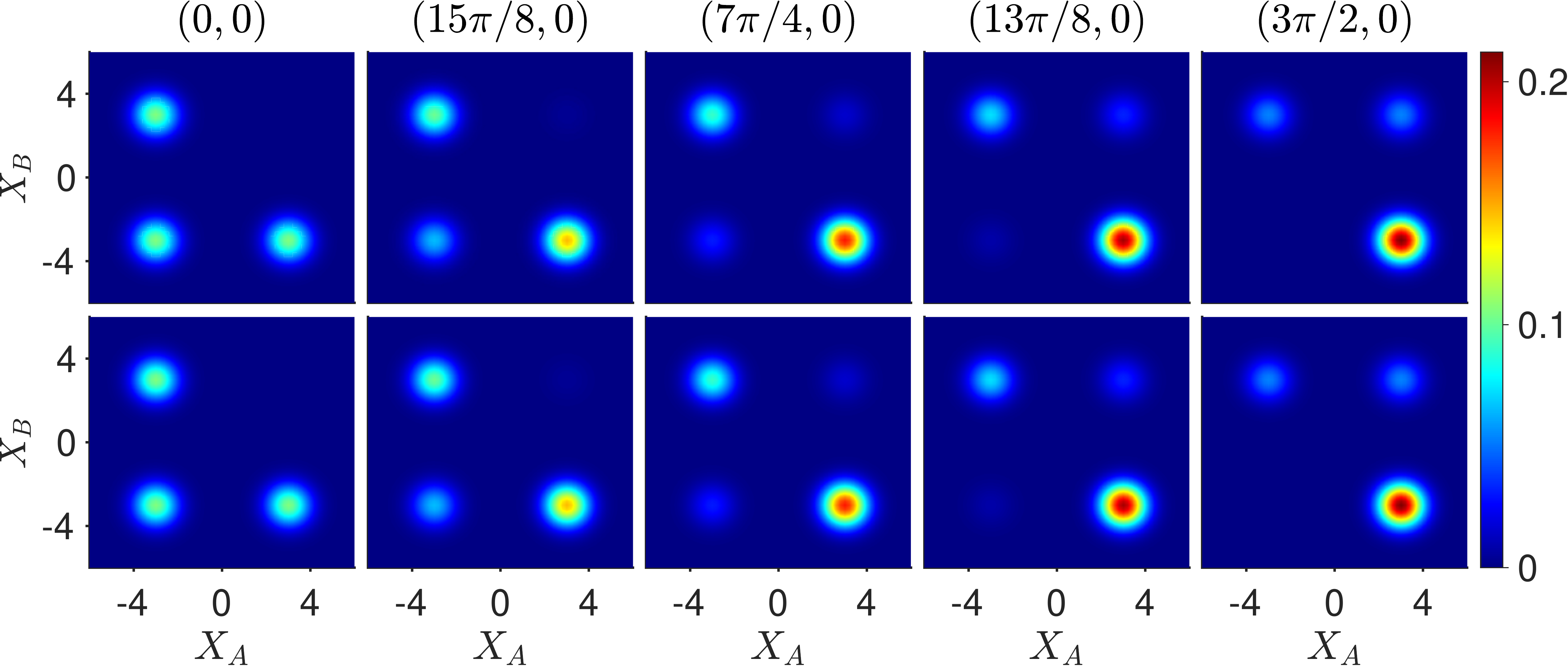}
\par\end{centering}

\caption{\textcolor{brown}{}Dynamics for the single unitary rotation $U$
associated with the joint measurements of $S_{y}^{A}$ and $S_{z}^{B}$
by the superobservers is indistinguishable from that of $\rho_{mix,B}$:
The notation is as for Figure \ref{fig:q-function-catstatezz-yz-1}.\textcolor{red}{{}
}Top sequence: The system begins in $|\psi_{zz}\rangle_{FR}$ (Eq.
(\ref{eq:ini})), and evolves according to $U_{A}^{-1}$ to the state
$|\psi_{yz}\rangle_{FR}$ (Eq. (\ref{eq:fryz})). Lower sequence:
The system begins in $\rho_{mix,B}$ and evolves as for the top plots.
The top and lower sequences are identical, which indicates consistency
with weak macroscopic realism (refer text).\textcolor{blue}{\label{fig:q-function-catstatezz-yz-1-1}}\textcolor{red}{}\textcolor{blue}{}\textcolor{red}{}}
\end{figure}

We first examine $\langle S_{y}^{A}S_{z}^{B}\rangle$. Here, the superobserver
$A$ would apply the unitary rotation $U_{A}^{-1}(t_{a})$. The dynamics
to create the state $|\psi_{yz}\rangle_{FR}$ is given by $U_{A}^{-1}(t_{a})|\psi_{zz}\rangle_{FR}$.
 The evolution is pictured as the top sequence of snapshots of
Figure \ref{fig:q-function-catstatezz-yz-1}. After an interaction
time $t_{a}=-\pi/2\equiv3\pi/2$, the state is $|\psi_{yz}\rangle_{FR}$,
for which the $Q$ function is\textcolor{blue}{}\textcolor{red}{}
\begin{eqnarray}
Q_{yz} & = & \frac{e^{-|\alpha|^{2}-X_{A}^{2}-P_{A}^{2}-|\beta|^{2}-X_{B}^{2}-P_{B}^{2}}}{6\pi^{2}}\Bigl\{4e^{2\left[\alpha X_{A}-\beta X_{B}\right]}\nonumber \\
 &  & +4\cos(2\left[\alpha P_{A}-\beta P_{B}\right])-4e^{2\alpha X_{A}}\sin(2\beta P_{B})\nonumber \\
 &  & +2e^{2\beta X_{B}}\cosh(2\alpha X_{A})-2e^{2\beta X_{B}}\sin(2\alpha P_{A})\Bigl\}.\nonumber \\
\label{eq:qyz}
\end{eqnarray}
The marginal function is
\begin{align}
Q_{yz}(X_{A},X_{B}) & =\frac{e^{-|\alpha|^{2}-X_{A}^{2}-|\beta|^{2}-X_{B}^{2}}}{3\pi}\Bigl\{2e^{2\left[\alpha X_{A}-\beta X_{B}\right]}\nonumber \\
 & {\color{black}+2e^{-\alpha^{2}-\beta^{2}}+e^{2\beta X_{B}}\cosh(2\alpha X_{A})\Bigl\}}\label{eq:Qmarg2}
\end{align}
as plotted in Figure \ref{fig:q-function-catstatezz-yz-1-1}. Including
the treatment of the final coupling to the superobservers' meters
$M$, as above, and then taking $\gamma$ large, we obtain for the
inferred measured amplitudes
\begin{align}
Q_{yz}(X_{A},X_{B}) & =\frac{e^{-|\alpha|^{2}-X_{A}^{2}-|\beta|^{2}-X_{B}^{2}}}{3\pi}\Bigl\{2e^{2\left[\alpha X_{A}-\beta X_{B}\right]}\nonumber \\
 & {\color{black}+e^{2\beta X_{B}}\cosh(2\alpha X_{A})\Bigl\}}.\label{eq:Qmarg3}
\end{align}
This agrees with that marginal (\ref{eq:Qmarg2}) derived directly
from the cat state where $\alpha=\beta$ is large, which is the case
of interest.

\begin{figure}[t]
\begin{centering}
\includegraphics[width=1\columnwidth]{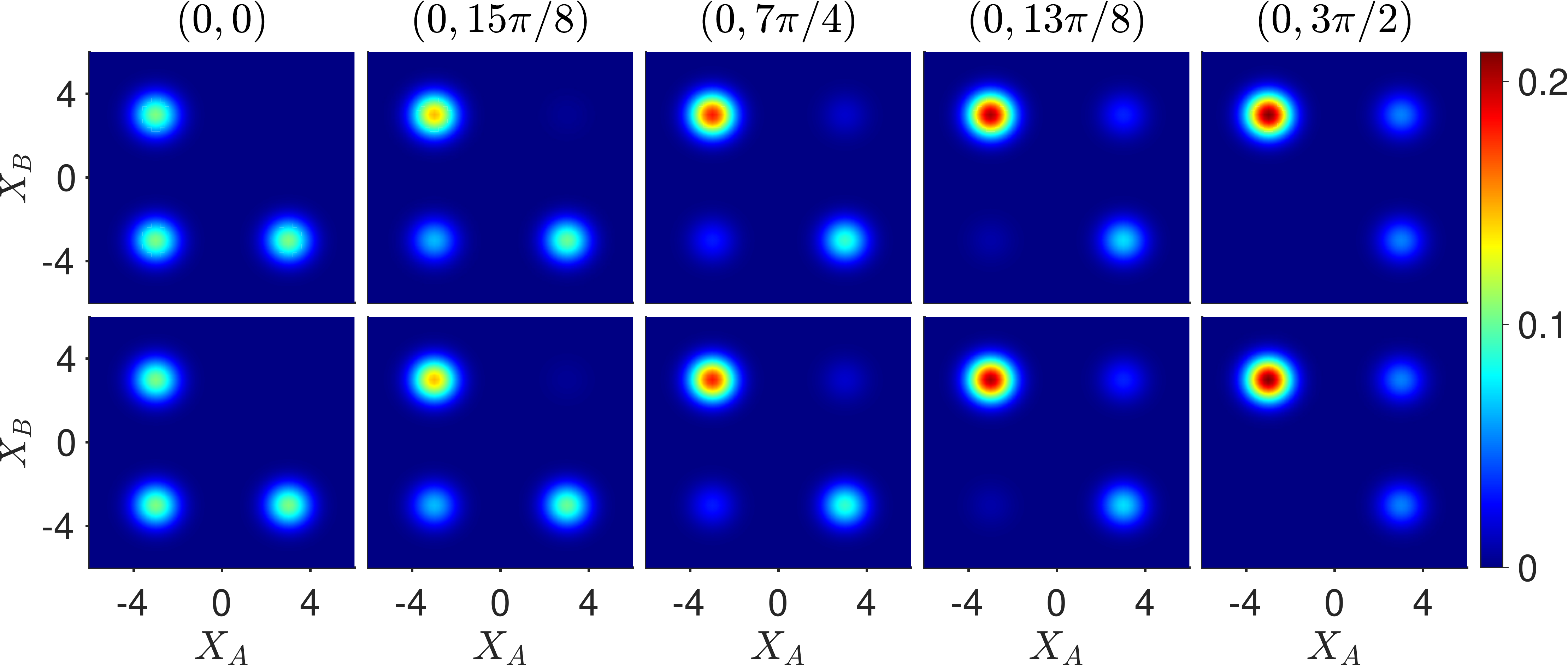}
\par\end{centering}
\caption{Dynamics for the single unitary rotation $U$ associated with the
joint measurements of $S_{z}^{A}$ and $S_{y}^{B}$ by the superobservers
is indistinguishable from that of $\rho_{mix,A}$, where Friend $B$'s
measurement is not reversed: The notation is as for Figure \ref{fig:q-function-catstatezz-yz-1}.
Top sequence: The system begins in $|\psi_{zz}\rangle_{FR}$ (Eq.
(\ref{eq:ini})) and evolves according to $U_{B}^{-1}$ to the state
$|\psi_{zy}\rangle_{FR}$ (Eq. (\ref{eq:catzy})). Lower sequence:
The system begins in $\rho_{mix,A}$ and evolves according to $U_{B}^{-1}$
as for the top plots. The two sequences are indistinguishable, indicating
consistency with weak macroscopic realism (refer text).\label{fig:q-function-catstatezz-zy-1-1-1}\textcolor{red}{}}
\end{figure}

Similarly, after the appropriate reversal, the measurement $S_{y}$
at $B$ requires the evolution $U_{B}^{-1}(t_{b})|\psi_{zz}\rangle_{FR}$,
which gives after a time $t_{b}=-\pi/2\equiv3\pi/2$, the state $|\psi_{zy}\rangle_{FR}$.
The dynamics for this measurement is plotted in Figure \ref{fig:q-function-catstatezz-zy-1-1-1}.\textcolor{red}{}

\begin{figure}[t]
\begin{centering}
\par\end{centering}
\begin{centering}
\includegraphics[width=1\columnwidth]{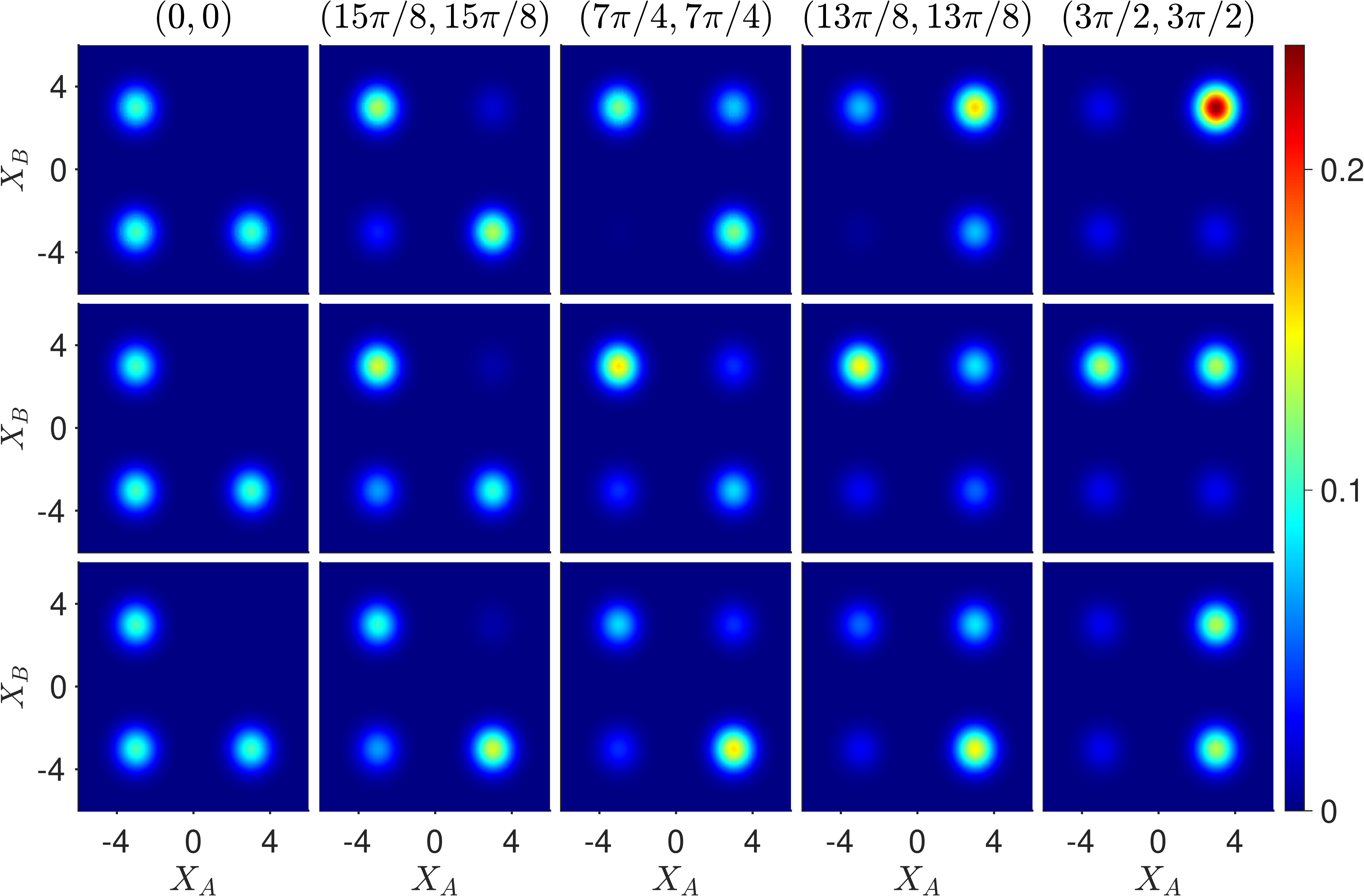}
\par\end{centering}
\begin{centering}
\par\end{centering}
\caption{\textcolor{brown}{}Dynamics for the two unitary rotations $U$ associated
with the joint measurements of $S_{y}^{A}$ and $S_{y}^{B}$ by the
superobservers. The notation is as for Figure \ref{fig:q-function-catstatezz-yz-1}.
Top sequence: The system begins in $|\psi_{zz}\rangle_{FR}$ (Eq.
(\ref{eq:ini})), and evolves according to $U_{A}^{-1}$ and $U_{B}^{-1}$
to the state $|\psi_{yy}\rangle_{FR}$ (Eq. (\ref{eq:cat-yy})). Centre
(lower) sequence: The system begins in $\rho_{mix,A}$ ($\rho_{mix,B}$)
 and evolves as for the top plots. \label{fig:q-function-catstatezz-yy-1}\textcolor{red}{}The
results for $|\psi_{zz}\rangle_{FR}$ diverge macroscopically from
those of the mixed states over the dynamics, leading to a violation
of the Wigner-Bell inequality for $|\psi_{zz}\rangle_{FR}$. This
is not inconsistent with wMR: The wMR premise allows a failure of
the macroscopic Bell-Locality premise where there are two rotations
(refer Section V.B.2).}
\end{figure}

After the rotations to measure $S_{y}$ at both sites, the system
is described by $U_{A}^{-1}U_{B}^{-1}|\psi_{zz}\rangle_{FR}$, as
given by $|\psi_{yy}\rangle$ (Eq. (\ref{eq:cat-yy})). \textcolor{blue}{}The
dynamics of these measurements in terms of the $Q$ function is plotted
in Figure \ref{fig:q-function-catstatezz-yy-1}.\textcolor{blue}{}\textcolor{red}{}\textcolor{red}{}

\subsubsection{Comparison with the classical mixture $\rho_{zz}$: the perspective
of the Friends}

The mixture $\rho_{zz}$ (Eq. (\ref{eq:mix-2})) is the state formed
from the perspective of the Friends, if the two Friends have both
measured $\sigma_{z}$ at their locations e.g. by coupling to the
meter. The $\rho_{zz}$ describes the statistical state of the system
\emph{conditioned} on \emph{both} the Friends' outcomes for their
measurements of $\sigma_{z}$. The state is conditioned on the outcome
$+1$ or $-1$ for the spins $S_{z}$ of the meter modes, denoted
$Am$ and $Bm$ in (\ref{eq:ini-2-1}), also found by integration
of the full $Q$ function over the system and meter variables as explained
in Section VI.A, to derive the result (\ref{eq:Qmarg_4mode-1-1-1-1}).
Each meter mode is coupled to the system, so that the systems themselves
can continue to evolve, conditioned on the outcome of the measurement
on the meter. This evolution is given by that of $\rho_{zz}$.

The dynamics for $\rho_{zz}$ is plotted in Figure \ref{fig:q-function-catstatezz-yz-1},
below the dynamics for the superposition state $|\psi_{zz}\rangle_{FR}$.
We see that the macroscopic difference between the predictions \emph{emerges
over the timescales of the unitary interaction} $U$ responsible for
the measurement settings.

The state $\rho_{zz}$ is consistent with a model for which there
is a definite predetermined outcome $\lambda_{z}^{A}$ and $\lambda_{z}^{B}$
for $\sigma_{z}$ in each Lab. The evolution for the system prepared
in $\rho_{zz}$ is given as $\rho(t_{a},t_{b})=U_{A}^{-1}U_{B}^{-1}\rho_{zz}U_{A}U_{B}$,
which satisfies a local realistic theory of the type considered by
Bell, and as such does not violate the Bell-Wigner inequality. Hence,
the paradox is not realised by $\rho_{zz}$. The superobservers must
reverse the Friends' measurements in order to restore the state $|\psi_{zz}\rangle_{FR}$.
We note the violation of the Bell-Wigner inequality can be inferred,
by performing measurements directly on $|\psi_{zz}\rangle_{FR}$,
based on the assumption that a measurement of $S_{z}$ made by the
superobservers would yield the same value as that of the Friends.
It is clear that the predictions and dynamics displayed by the entangled
state $|\psi_{zz}\rangle_{FR}$ are not compatible with the mixed
state $\rho_{zz}$ (nor any state giving consistency with a local
realistic theory, since such a state would not violate (\ref{eq:Bell-CHSH})).

\subsubsection{Comparison with partial mixtures: conditioning on one Friend's measurement}

We now compare the evolution of $|\psi_{zz}\rangle_{FR}$ with that
of partial mixtures obtained by conditioning on the outcomes of one
Friend. This enables demonstration of the consistency with wMR, using
the Result 6.B. (1).

First, we consider the dynamics for the measurements of $S_{z}^{B}$
and $S_{y}^{A}$, on the system prepared in the state $|\psi_{zz}\rangle_{FR}$.
This dynamics creates $|\psi_{yz}\rangle_{FR}$. We compare with
the mixture created if the spin measurement $\sigma_{z}^{B}$ made
by the Friend in $L_{B}$ is not reversed. The Friend has coupled
the meter $Bm$ to the system $B$, as in (\ref{eq:ini-2-1}), and
conditions all future measurements on the outcome for the meter being
$+1$ or $-1$. The density operator for the combined system after
a measurement of spin $\sigma_{z}$ at $B$ is the partial mixture
\begin{eqnarray}
\rho_{mix,B} & = & \frac{1}{3}(|-\alpha\rangle\langle-\alpha|)(|\beta\rangle\langle\beta|)\nonumber \\
 &  & +\frac{1}{3}\Bigl\{(|\alpha\rangle+i|-\alpha\rangle)(\langle\alpha|-i\langle-\alpha|)\Bigl\}(|-\beta\rangle\langle-\beta|).\nonumber \\
\label{eq:semib}
\end{eqnarray}
\textcolor{blue}{}Now we consider that a measurement $S_{y}^{A}$
is made on system $A$. This implies (after the appropriate reversal)
the evolution according to $U_{A}^{-1}(t_{a})$, which is given in
Figure \ref{fig:q-function-catstatezz-yz-1-1} for both $|\psi_{zz}\rangle_{FR}$
and $\rho_{mix,B}$. The evolution of $\rho_{mix,B}$ is indistinguishable
from that of $|\psi_{zz}\rangle_{FR}$.\textcolor{blue}{}

We next consider the dynamics associated with the measurements of
$S_{z}^{A}$ and $S_{y}^{B}$ on the system prepared in $|\psi_{zz}\rangle_{FR}$,
which creates $|\psi_{zy}\rangle_{FR}$. For comparison, we also
consider that the Friend at Lab A performs the measurement $\sigma_{z}^{A}$.
The conditioning on the outcomes for the Friend's measurement leaves
the systems in the mixture
\begin{eqnarray}
\rho_{mix,A} & = & \frac{1}{3}(|\alpha\rangle\langle\alpha|)(|-\beta\rangle\langle-\beta|)\nonumber \\
 &  & +\frac{1}{3}(|-\alpha\rangle\langle-\alpha|)\Bigl\{\bigl(|\beta\rangle+i|-\beta\rangle\bigl)\bigr(\langle\beta|-i\langle-\beta|\bigl)\Bigl\}.\nonumber \\
\label{eq:semia}
\end{eqnarray}
The dynamics associated with the measurement of $S_{y}^{B}$ involves
the system evolving according to $U_{B}^{-1}(t_{b})$ is given in
Figure \ref{fig:q-function-catstatezz-zy-1-1-1}. We see that the
evolution for systems prepared in $|\psi_{zz}\rangle_{FR}$ is indistinguishable
from that of systems prepared in $\rho_{mix,A}$.\textcolor{blue}{}

Finally, we consider the dynamics where measurements of $S_{y}^{A}$
and $S_{y}^{B}$ are made on the system prepared in the state $|\psi_{zz}\rangle_{FR}$.
Here, two local unitary rotations $U_{A}^{-1}(t_{a})$ and $U_{B}^{-1}(t_{b})$
are applied, to create the state $|\psi_{yy}\rangle_{FR}$, prepared
in the pointer bases for the measurements $S_{y}^{A}$ and $S_{y}^{B}$.
We compare this evolution with that of the system prepared in $\rho_{mix,A}$,
or $\rho_{mix,B}$ (Figure \ref{fig:q-function-catstatezz-yy-1}).
The evolution is \emph{macroscopically different} in each case.

\subsubsection{Consistency with the weak macroscopic realism model}

The predictions of quantum mechanics as given in Figures \ref{fig:q-function-catstatezz-yz-1-1}-\ref{fig:q-function-catstatezz-yy-1}
reveal consistency with weak macroscopic realism (wMR).  First, consider
measurement of $\langle S_{y}^{A}S_{z}^{B}\rangle$. The evolution
of $|\psi_{zz}\rangle_{FR}$ shown in Figure \ref{fig:q-function-catstatezz-yz-1-1}
is indistinguishable from that of $\rho_{mix,B}$. Hence, from Result
6.B.1, those results show consistency with wMR. After the interaction
$U_{A}^{-1}$ in Lab $A$, the system is prepared in the appropriate
basis, so that the final pointer measurement $S_{y}^{A}$ can be made
by the superobserver. Hence, by Result 6.A, this state is also consistent
with wMR. Similarly, prior to the dynamics $U_{A}^{-1}$ portrayed
in the Figure \ref{fig:q-function-catstatezz-yz-1-1}, the superobservers
perform the reversal of the Friends' measurements. This does not change
the preparation basis and, as argued in Section VI.A, a wMR model
exists  in which the pointer value $\lambda_{z}^{B,S}=\lambda_{z}^{B}$
is unchanged. We therefore conclude consistency with wMR.

The same arguments apply to the measurement of $\langle S_{z}^{A}S_{y}^{B}\rangle$.
The evolution of $|\psi_{zz}\rangle_{FR}$ shown in Figure \ref{fig:q-function-catstatezz-zy-1-1-1}
is indistinguishable from that of $\rho_{mix,A}$. Hence, from Result
6.B.1, those results show consistency with wMR.

We claim a particular wMR model exists that replicates the quantum
predictions of the $|\psi_{zz}\rangle_{FR}$, for the moments $\langle S_{z}^{A}S_{z}^{B}\rangle$,
$\langle S_{z}^{A}S_{y}^{B}\rangle$ and $\langle S_{y}^{A}S_{z}^{B}\rangle$.
At first glance the model seems not consistent, since for the different
moments, we consider different mixed states ($\rho_{zz}$, $\rho_{mix,A}$
and $\rho_{mix,B}$) which are not compatible. In fact, we propose
a more complete wMR model which includes local operations made by
the superobservers. In the model, the system begins in $\rho_{zz}$
at the time $t_{2}$, and consistent with that mixed state, the results
for both Friends' measurements of $\sigma_{z}$ are known to the superobservers.
Where the superobserver $A$ measures $S_{y}$ by applying $U_{A}^{-1}$,
then in the model the superobserver $A$ first operates locally in
Lab $A$ so that the overall system in $\rho_{zz}$ is transformed
into $\rho_{mix,B}$. This local operation does not change the value
of $\lambda_{z}^{B}$ in the model. Similarly, if superobserver $B$
measures $S_{y}$, then in the wMR model, a local operation is performed
to change $\rho_{zz}$ into $\rho_{mix,A}$. For this model, wMR holds
throughout the dynamics.

In summary, we have shown consistency of the quantum predictions of
the Wigner friends paradoxes with both the wMR assertions. This was
done using a particular wMR model, and showing compatibility with
the predictions for $\langle S_{z}^{A}S_{z}^{B}\rangle$, $\langle S_{z}^{A}S_{y}^{B}\rangle$
and $\langle S_{y}^{A}S_{z}^{B}\rangle$. However, the particular
wMR model used is a Bell-local realistic one, and does not describe
the quantum dynamics for the measurements of $S_{y}$ at both Labs
i.e. where there are two rotations $U$, one at $A$ and one at $B$.
 We see from Section V however that this does not imply the quantum
predictions are inconsistent with the premise of wMR. The two assertions
of wMR are not testable where there are unitary rotations at \emph{both}
sites (refer Figure \ref{fig:q-function-catstatezz-yy-1}), since
both systems shift to a new pointer basis, so that the former pointer
value $\lambda$ according to wMR no longer applies.

\section{Conclusion and discussion}

The motivation of this paper is to present a mapping between the microscopic
Wigner friend paradoxes involving spin qubits and macroscopic versions
involving macroscopically distinct spin states. In Section III, we
provide such a mapping, where the macroscopically distinct states
are two coherent states, $|\alpha\rangle$ and $|-\alpha\rangle$,
$\alpha\rightarrow\infty.$ Unitary rotations $U$ determine which
spin component is to measured and in a microscopic spin experiment
correspond to Stern-Gerlach or polarizer-beam-splitter analyzers.
In the macroscopic set-up, the $U$ are realised with nonlinear interactions.

The mapping motivates us to seek an interpretation for the paradoxes
where macroscopic realism can be upheld. The extended Wigner-Friend
paradoxes are based on very reasonable assumptions, including that
of Locality, defined by Bell. We show in Section IV that the realisations
of Brukner's Bell-Wigner friend and the Frauchiger-Renner paradoxes
each imply falsification of deterministic macroscopic (local) realism.

Motivated by that, we consider in Section V a more minimal definition
of macroscopic realism, called weak macroscopic realism (wMR), which
assigns realism to the system as it exists after the unitary rotation
$U$ that determines the measurement setting. This establishes a predetermined
value for the outcome of the pointer measurement that is to follow.
The premise of wMR also establishes a locality \emph{for this value}:
the value is not affected by events at a spacelike-separated Lab.
Careful examination shows that wMR does \emph{not} imply a full Locality
of the type postulated by Bell, which defines a realism for the system
as it exists prior to the unitary dynamics $U$. In the remaining
Sections VI-VII, we prove several Results, which confirm that the
predictions of quantum mechanics for the paradoxes are consistent
with wMR.

A feature of wMR is the definition of realism in a contextual sense.
In the wMR model, the quantum state is not defined completely until
the basis of preparation is specified. The basis of preparation for
a particular state at a given time $t$ is defined as the basis such
that the spin can be measured without a further unitary rotation $U$
that would give a change of measurement setting. The final measurement
involves a sequence of operations such as amplification and detection,
or coupling to meters. These operations are referred to as the final
pointer stage of the measurement.

It is possible to define a similar contextual realism for the microscopic
qubits. We refer to this as \emph{weak contextual realism} (wR or
wLR). The interpretations of the macroscopic paradoxes can be replicated
in the microscopic versions, since there is a mapping between the
two. The proofs of the Results in Sections IV-VI follow identically,
for the spin $1/2$ system. This implies failure of deterministic
local realism, and consistency with weak contextual realism. The violation
of the Bell-CHSH and Brukner's Bell-Wigner inequality is possible,
because wLR does \emph{not} imply the full Bell Locality assumption.

The results of this paper give insight into how the assumptions of
Bell's theorem may break down for quantum mechanics. We find the paradoxes
arise only where the measurement setting is changed at \emph{both}
sites. This implies two unitary rotations. The unitary dynamics has
been analyses using the $Q$ function. There is an effectively unobservable
(as $\alpha\rightarrow\infty$) difference between the Q function
of the macroscopic superposition and that of the corresponding mixture
(the mixture giving consistency with the Bell-CHSH inequalities).
The difference remains undetectable for the case where there is only
a single rotation, which allows a predetermination of one of the measurement
outcomes, so that wMR applies. However, where there are two rotations,
the functions for the states evolving from the macroscopic superposition
and the mixture become \emph{macroscopically} different. This leads
to macroscopic differences in the predictions, hence allowing the
macroscopic paradox. This is the strangest mathematical paradox,
because the final difference between the functions is macroscopic
in the limit $\alpha\rightarrow\infty$, precisely the limit where
the initial difference is increasingly negligible.

While we propose that wMR (and wLR) holds, we do not present a full
wMR model consistent with all the quantum predictions of the paradox.
The quantum predictions have been shown consistent with wMR only where
the postulate wMR applies, which means where there has been a final
measurement setting established. This leaves open the question of
whether wMR is truly compatible with quantum mechanics, or whether
it can be falsified. Arguments have been given elsewhere that there
is inconsistency between wMR and the completeness of (standard) quantum
mechanics \citep{ghz-cat,manushan-bell-cat-lg,s-cat}. This motivates
examination of alternative theories, or theories which may give a
more complete description of quantum mechanics (e.g. \citep{DrummondReid2020,bohm,hall-cworlds}),
for consistency with wMR.

Finally, we consider a possible experiment. The microscopic superposition
states can be mapped onto coherent-state superpositions using the
methods of \citep{cat-det-map,cat-bell-wang-1}. The unitary rotations
involving Kerr interactions have been realised in experiments creating
cat states \citep{cat-states-super-cond}. The experiments could also
be conducted using Greenberger-Horne-Zeilinger states and CNOT gates,
as in \citep{macro-realism-nori}.

\section*{Acknowledgements}

This research has been supported by the Australian Research Council
Discovery Project Grants schemes under Grant DP180102470 and DP190101480.
The authors also wish to thank NTT Research for their financial and
technical support.

\section{Appendix}

\subsection{Examples of realisation of the Friend's cat states}

In this paper, we give three examples of a Wigner's friend experiment
where the initial spin system is a macroscopic one. The second example
uses GHZ states and CNOT operations. Consider a large number of spin
$1/2$ qubits. For Lab $A$, we select a set of $N+1$ qubits, choosing
$|h\rangle=|\uparrow\rangle|\uparrow\rangle^{\otimes N}$ and $|t\rangle=|\downarrow\rangle|\downarrow\rangle^{\otimes N}$.
Similar macroscopic qubits can be selected for Lab $B$. The qubits
can be realised as orthogonally polarized photons in $N+1$ different
modes. The coupling to the meters in the Labs links the systems to
a larger set of $M$ qubits, so that $|H\rangle=|\uparrow\rangle|\uparrow\rangle^{\otimes N}|\uparrow\rangle^{\otimes M}$
and $|T\rangle=|\downarrow\rangle|\downarrow\rangle^{\otimes N}|\downarrow\rangle^{\otimes M}$.
The unitary rotation $U_{x}$ or $U_{y}$ can be realised using CNOT
gates. Suppose the system $A$ is created in $|H\rangle$. Then the
photon of the first mode is passed through a beam splitter or polarizer-type
interaction, to create
\begin{equation}
(|\uparrow\rangle+e^{i\varphi}|\downarrow\rangle)|\uparrow\rangle^{\otimes(N+M)}.\label{eq:statephase}
\end{equation}
 For each subsequent qubit, a CNOT operation is applied, which creates
the Greenberger-Horne-Zeilinger (GHZ) state \citep{ghz-amjp,mermin-inequality,omran-cats,ghz-1}\textcolor{red}{}
\begin{equation}
|\uparrow\rangle|\uparrow\rangle^{\otimes(N+M)}+e^{i\varphi}|\downarrow\rangle|\downarrow\rangle^{\otimes(N+M)}.\label{eq:macro-ghz}
\end{equation}
The inclusion of a phase shift $\varphi$ at the first mode transformation
allows either the $U_{x}$ or the $U_{y}$ to be realised. Such states
have been used to experimentally demonstrate failure of macrorealism
\citep{macro-realism-nori}, and have also been proposed for macroscopic
tests of GHZ and Bohm-Einstein-Podolsky-Rosen paradoxes \citep{ghz-cat},
as well as for testing macroscopic Bell inequalities \citep{manushan-bell-cat-lg}.

The third example uses two-mode states and nonlinear interactions.
The macroscopic qubits are two-mode number states, given by $|N\rangle_{1}|0\rangle_{2}$
and $|0\rangle_{1}|N\rangle_{2}$ where $|n\rangle_{i}$ is a number
state for the mode $i$. These states for large $N$ are macroscopically
distinct. The states were studied in \citep{macro-bell-lg} and \citep{ghz-cat},
where it was shown that a nonlinear interaction $H_{nl}$ can create
the macroscopic cat superposition according to $U_{x}$ and $U_{y}$
given by (\ref{eq:ransx}) and (\ref{eq:transy}) (where we put $|H\rangle=|N\rangle|0\rangle_{2}$
and $|T\rangle=|0\rangle_{1}|N\rangle_{2}$). The transformations
are not fully realised, but are sufficiently effective that violation
of Bell inequalities are predicted.

\subsection{Meter coupling}

Let us consider the qubit $c|\uparrow\rangle+d|\downarrow\rangle$,
where $c$ and $d$ are complex amplitudes. We couple the qubit system
to a field mode prepared initially in a coherent state $|\gamma_{0}\rangle$.
We consider the evolution under $H_{Am}$ where \citep{spin-meter,Ilo-Okeke-byrnes-meter,blais-meter-model,spin-coupling}
\begin{equation}
H_{Am}=\hbar G\sigma_{z}^{A}n_{c}^{A}.\label{eq:measurement-1}
\end{equation}
Here $\sigma_{z}^{A}$ is the Pauli spin operator for the qubit system
$A$, $n_{c}^{A}$ is the number operator for the meter mode $C$
in Lab A, and $G$ is a real constant. The solution for the final
entangled state is
\begin{eqnarray*}
|\psi\rangle_{out} & = & e^{-iH_{d}t/\hbar}(c|\uparrow\rangle|\gamma_{0}\rangle+d|\downarrow\rangle|\gamma_{0}\rangle)\\
 & = & c|\uparrow\rangle|\gamma\rangle+d|\downarrow\rangle|-\gamma\rangle
\end{eqnarray*}
where we select $Gt=\pi/2$ and $\gamma=-i\gamma_{0}$. 

\onecolumngrid

\subsection{Table of values for macroscopic realistic states}

\begin{tabular}{|c|c|c|c|c|c|c|c|c|c|}
\hline 
$\lambda_{zA}$ & $\lambda_{zB}$ & $\lambda_{yA}$ & $\lambda_{yB}$ & $P_{--|yy}$ & $P_{-+|yy}$ & $P_{+-|yy}$ & $P_{++|yy}$ & $P_{--|yz}$ & $P_{--|zy}$\tabularnewline
\hline 
\hline 
1 & 1 & 1 & 1 & 0 & 0 & 0 & 1 & 0 & 0\tabularnewline
\hline 
1 & 1 & 1 & -1 & 0 & 0 & 1 & 0 & 0 & 0\tabularnewline
\hline 
1 & 1 & -1 & 1 & 0 & 1 & 0 & 0 & 0 & 0\tabularnewline
\hline 
\textcolor{blue}{{*} 1} & \textcolor{blue}{{*}1} & \textcolor{blue}{{*}-1} & \textcolor{blue}{{*}-1} & \textcolor{blue}{{*}1} & \textcolor{blue}{{*}0} & \textcolor{blue}{{*}0} & \textcolor{blue}{{*}0} & \textcolor{blue}{{*}0} & \textcolor{blue}{{*}0}\tabularnewline
\hline 
1 & -1 & 1 & 1 & 0 & 0 & 0 & 1 & 0 & 0\tabularnewline
\hline 
1 & -1 & 1 & -1 & 0 & 0 & 1 & 0 & 0 & 0\tabularnewline
\hline 
1 & -1 & -1 & 1 & 0 & 1 & 0 & 0 & 1 & 0\tabularnewline
\hline 
1 & -1 & -1 & -1 & 1 & 0 & 0 & 0 & 1 & 0\tabularnewline
\hline 
-1 & 1 & 1 & 1 & 0 & 0 & 0 & 1 & 0 & 0\tabularnewline
\hline 
-1 & 1 & 1 & -1 & 0 & 0 & 1 & 0 & 0 & 1\tabularnewline
\hline 
-1 & 1 & -1 & 1 & 0 & 1 & 0 & 0 & 0 & 0\tabularnewline
\hline 
-1 & 1 & -1 & -1 & 1 & 0 & 0 & 0 & 0 & 1\tabularnewline
\hline 
-1 & -1 & 1 & 1 & 0 & 0 & 0 & 1 & 0 & 0\tabularnewline
\hline 
-1 & -1 & 1 & -1 & 0 & 0 & 1 & 0 & 0 & 1\tabularnewline
\hline 
-1 & -1 & -1 & 1 & 0 & 1 & 0 & 0 & 1 & 0\tabularnewline
\hline 
-1 & -1 & -1 & -1 & 1 & 0 & 0 & 0 & 1 & 1\tabularnewline
\hline 
\end{tabular}

\vspace{2cm}

\twocolumngrid

\end{document}